# Identifying Planetary Nebulae Candidates from IPHAS and VPHAS+ Surveys via YOLO Deep Learning Models

Titouan Sérandour[1,2], Anne-Laure Méalier[3], Bruno Colicchio[1], Dominique Daniel[4], Philippe Gilberton[4], Philippe Oros[4], Thomas Boch[5], and Djaffar Ould Abdeslam[1]
AIP-PN

[1] Institut de Recherche en Informatique, Mathématiques, Automatique et Signal (IRIMAS), Université de Haute-Alsace, 61 Rue Albert Camus, 68093 Mulhouse, France; name.surname@uha.fr
[2] Centrale Méditerranée, 38 Rue Frédéric Joliot-Curie, 13451 Marseille, France
[3] Aix Marseille Univ, CNRS, Centrale Med, Institut Fresnel, Marseille, France
[4] Astro Images Processing (AIP), France
[5] Université de Strasbourg, CNRS, Observatoire astronomique de Strasbourg, UMR 7550, F-67000 Strasbourg, France
*Received 2026 February 10; revised 2026 May 26; accepted 2026 June 6; published 2026 September 29*

## Abstract

We present a deep learning detection method capable of effectively identifying planetary nebulae (PNe) on IPHAS and VPHAS+ surveys. Our detection pipeline is based on the YOLO architecture and is adaptable to other datasets. Using the HiPS system, we preprocess and augment the *r*, *i*, and H$\alpha$ bands to produce two red, green, and blue (RGB) image datasets: (1) 640 × 640 px images at full resolution, and (2) 1 280 × 1 280 px images designed to capture extended objects up to 0.2 deg. Images of ∼2600 PNe are labelled using the HASH database. The loss function and evaluation metrics are adapted to the diffuse and low-contrast nature of PNe; a very low intersection over union (IoU) threshold (1%) is used. A *k*-fold cross validation shows that the YOLOv12m model achieves an average precision of 0.979 on the first dataset. At a confidence threshold of 0.15, it achieves a precision rate of 98.7% and a recall rate of 97.4%. We deployed our system (on 1280 × 1280 images) on the full surveys' coverage (∼3755 deg$^2$), and present a preliminary analysis of the detection results. The system is robust against artifacts such as CCD corner defects and HiPS tile joints. Our convolutional-neural-network-based model achieved excellent results on the test set, comparable to—or exceeding—that of a recent Swin Transformer approach. However, a challenge persists when applying deep learning models to extensive and densely populated Galactic fields: the development of an automated system that can effectively discriminate PNe from mimics that human experts are able to distinguish (e.g., H II regions). We outline possible improvements for future work.

## 1. Introduction

A planetary nebula (PN) is a celestial object consisting of an expanding, glowing envelope ejected from a low- or intermediate-mass star (a star of around 1–8 solar masses) late in its life. The elevated temperature of the central star, from which the gas originated, causes the ejected material to become ionized, making it visible as a PN. Identifying PNe in astronomical photometric surveys is a complex problem that has been tackled using both visual inspection and automated methods over the past decades. The Galactic plane—which contains most PNe observable from Earth—is a vast region of the sky, crowded with celestial sources, some of which may be mistaken for PNe or hide them. In addition, PNe exhibit a wide range of morphologies and can be very faint. Consequently, their detection and classification remain challenging for both manual and automated approaches.

Recent advances in wide-field, narrowband imaging surveys, especially those centered on the H$\alpha$ emission line, have transformed the discovery of PNe. Over the last decades, deep and high-resolution H$\alpha$ surveys such as the SuperCOSMOS H$\alpha$ Survey (SHS) and the Isaac Newton Telescope (INT) Photometric H$\alpha$ Survey of the Northern Galactic Plane (IPHAS) have dramatically expanded our catalog of Galactic and Magellanic Cloud PNe, more than doubling the number found in the previous centuries. A large number of them have been discovered via the visual scrutiny of IPHAS, by both professionals (L. Sabin et al. 2014, 2022) and amateurs (P. Le Dû et al. 2022).

The introduction of artificial neural networks to this field can be traced to M. Faúndez-Abans et al. (1996), who employed them to classify PNe based on their chemical composition. More recently, S. Akras et al. (2019) and G. Liberato et al. (2023) used classification tree algorithms to find infrared criteria (applied on the Two Micron All Sky Survey and AllWISE photometric data) that can distinguish compact PNe from their mimics in the photometric catalogs of IPHAS and of the Very Large Telescope (VLT) Survey Telescope (VST) Photometric H$\alpha$ Survey of the Southern Galactic Plane and Bulge (VPHAS+), as well as in the list of PNe candidates published by K. Viironen et al. (2009a, 2009b). D. N. F. Awang Iskandar et al. (2020) used transfer learning to test various deep learning networks, in particular DenseNet201, to decide whether an object is a PN and to determine its morphological class, using images gathered in the Hong Kong/AAO/Strasbourg H$\alpha$ (HASH) PN database.

Recently, R. Sun et al. (2024) developed a deep learning tool to identify PN candidates in VPHAS+, which had not yet undergone extensive manual, systematic searching. They used a model based on the Mask R-CNN architecture, replacing the ResNet feature extraction network with a Swin Transformer network. After spectroscopic follow-up observations of top-quality candidates, they identified 22 new True, Likely, and Possible PNe; results were presented in a second paper by Y. Li et al. (2024).

In our work, we adopted the You Only Look Once (YOLO) framework, a deep learning architecture based on convolutional neural networks (CNNs) and designed for real-time object detection. The most recent version, YOLOv12, incorporates an area attention module along with a residual efficient layer aggregation network (R-ELAN), enhancing its feature representation and detection performance. The use of YOLO-based deep learning models in astronomy has recently been explored successfully in tasks such as the detection of blue horizontal-branch stars (Y. He et al. 2023), galaxy clusters (K. Grishin et al. 2023), galaxies (R. González et al. 2018; D. Cornu et al. 2024), deep-sky objects (O. Parisot & M. Jaziri 2024), strong gravitational lenses (Y. Liu et al. 2025), and small-size dwarf galaxy candidates (H. Yang et al. 2025). See also the work of L. T. Ramos & F. Rivas-Echeverría (2025). To the best of our knowledge, the present study represents the first application of YOLO to the specific task of detecting PNe candidates.

Taking advantage of the high degree of homogeneity between IPHAS (J. E. Drew et al. 2005) and VPHAS+ (J. E. Drew et al. 2014) data, we build a completely unified pipeline for preprocessing and augmenting data from these two surveys, based in particular on the Hierarchical Progressive Surveys (HiPS) hierarchical tiling system developed at the Strasbourg Astronomical Data Center (CDS) (P. Fernique et al. 2017; T. Boch et al. 2020). We believe that this simplified approach is also highly generalizable to other applications of machine learning techniques to astronomy, considering the large number of HiPS available online. Thanks to this shared pipeline, we have a larger set of images to train our models. We combine the $r$, $i$, and H$\alpha$ spectral bands to construct two sets of red, green, and blue (RGB) images corresponding to two different fields of view, in order to target compact PNe as well as more extended, rare PNe with apparent diameters of up to $0\overset{\circ}{.}2$. We also take the diffuse and cloudy morphology of PNe into account in YOLO's loss function weighting as well as in the model evaluation methodology.

This work is part of the AIP-PN project (see D. Daniel 2024), conducted by a group of French amateur astrophotographers from the Astro Images Processing (AIP) association, dedicated to PNe research and imaging. AIP is one of the major astrophotography associations worldwide, dedicated to the pursuit of excellence in capturing and processing astronomical images. The first prediction results from the large-scale deployment of our model on IPHAS and VPHAS+ data are currently being analyzed by members of the project. The most promising candidates will be selected for a follow-up program of photometric and spectroscopic observations, to be carried out in the coming months by a larger team of AIP members. This will provide additional evidence to confirm the PN nature of the candidates. This collaboration between amateurs and professionals aims both to advance scientific knowledge of PNe and to produce visually inspiring astronomical images.

This paper is organized as follows. In Section 2, we introduce the problem of identifying PNe, their main properties, as well as the origin and specifications of the data we use. Section 3 is devoted to the exposition of our methodology, from data preprocessing, augmentation, and annotation to model performance evaluation metrics. In Section 4, we demonstrate the excellent performance achieved by YOLO models in the detection of PNe in IPHAS and VPHAS+ surveys and expose preliminary results of large-scale deployment.

*Open access to our code.* Our code is accessible from the AIP-PN-YOLO GitHub repository,[6] and the static version used in this work is available in Zenodo (T. Sérandour & A.-L. Mealier 2026).

## 2. PNe and Galactic Surveys Properties

### 2.1. PNe and Their Identification

PNe represent a brief (typically 25,000 ± 5000 yr) yet crucial evolutionary phase for intermediate-mass stars, tracing mass loss, chemical enrichment, and Galactic evolution. They also serve as key kinematic tools and distance indicators. The estimated number of Galactic PNe ranges from approximately 6600 to 45,000, which is much higher than the number of known PNe (less than 4000) and seems to suggest that many discoveries are still to come. For further information, we refer to D. J. Frew & Q. A. Parker (2010), Q. A. Parker (2022), K. B. Kwitter & R. B. C. Henry (2022) and the references therein. Besides their scientific interest, PNe are also beautiful and photogenic deep-sky objects whose observation is highly prized by amateur astronomers and astrophotographers.

In the past, a significant number of celestial objects were mistakenly considered to be PNe and were later identified as contaminants (PNe mimics) such as compact H II regions. This demonstrated the need to develop a reliable diagnostic method. Such a method was developed by the HASH database team (see Q. A. Parker 2022). A key to PN identification is spectroscopy. Though PNe spectra exhibit variability (in particular, along their evolutionary stage), typical emission lines include doubly ionized oxygen [O III] and hydrogen-alpha (H$\alpha$).

### 2.2. HASH and PlanetaryNebulae.net Catalogs

More than 3500 PNe are already known and cataloged. We rely on the HASH PN catalog, a comprehensive structured query language (SQL) online database[7] maintained by Andreas Ritter and Quentin Parker (Q. A. Parker et al. 2016). This database gathers information from various catalogs, notably *Li-Parker-1* (R. Sun et al. 2024), *2014MNRAS.443.3388S* (L. Sabin et al. 2014), and *FrenchAmateurs*. The PlanetaryNebulae.net website,[8] maintained by Pascal Le Dû and Thomas Petit, also contains a remarkable catalog of PNe, candidates, and related objects.

The HASH database brings together imaging, spectroscopic, and other data for all known galactic and Magellanic Cloud PNe, including their astronomical coordinates, angular diameters (expressed in arcseconds), and morphology types,

6 URL: https://github.com/CentraleDigitaleLab/AIP-PN-YOLO
7 URL: http://www.hashpn.space/ (registration required).
8 URL: https://planetarynebulae.net/

**Table 1**
Key Properties of IPHAS and VPHAS+ Surveys

| | IPHAS | VPHAS+ |
|---|---|---|
| Observation period | 2003.08—2012.11 | 2011.12—2018.08 |
| Approximate coverage | $\|b\| < 5°$, $29° < l < 215°$ | $\|b\| < 5°$, $210° < l < 40°$ |
| | | or $\|b\| < 10°$, $350° < l < 10°$ |
| Area (deg$^2$) | ~1800 | ~2000 |
| Telescope name | INT | VST |
| Telescope diameter (m) | 2.54 | 2.65 |
| Filters | H$\alpha$, $r$, $i$ | H$\alpha$, $u$, $g$, $r$, $i$ |
| Depth (mag) | ~21 ($r$), 20 ($i$), 20 (H$\alpha$) | ~20 (all), 22 (g) |
| Camera | Wide Field Camera | OmegaCAM |
| Angular resolution (arcsec/px) | 0.33 | 0.21 |

**Note.** Letters $l$ and $b$ stand for Galactic latitude and longitude, respectively.

though some entries lack these last two pieces of information. PNe in the database are classified with respect to their status: *True* (3 345 entries), *Likely* (462), or *Possible* (888), and we refer to them as "known" or "cataloged" PNe, since HASH is currently the most comprehensive PNe catalog. Our focus is on Galactic PNe. The database also includes objects that are not PNe, primarily sources that were previously considered to be PNe or PNe candidates.

### *2.3. IPHAS and VPHAS+ Galactic Surveys*

We use the data obtained as part of IPHAS DR2[9] (J. E. Drew et al. 2005; G. Barentsen et al. 2014) and VPHAS+ DR4[10] (J. E. Drew et al. 2014). The IPHAS survey was carried out by the INT, located in the Canary Islands, whereas the VPHAS+ survey was carried out by the VST, in Chile. See Table 1 for more detailed specifications; note that the surveys' footprints do overlap.

Both cameras consist of a mosaic of 2048 × 4100 px charge-coupled devices (CCDs). The observation process includes multiple exposures at different offset pointings (to make sure that only a very small portion of the survey footprint is lost because of the gaps between the CCDs in the camera mosaic) and with different passband filters placed in front of the camera: H$\alpha$ narrow band, as well as Sloan $u$, $g$, $r$, and $i$ broad bands ($r$ includes H$\alpha$ in its bandpass). The [O III] emission line is contained in the $g$ band.

The IPHAS survey presents a number of data quality issues, generating artifacts potentially interfering with object detection. Figure 10 shows illustrative examples of the robustness of our detection model to artifacts such as CCD corner defects.

### *2.4. HiPS*

In our work, we chose to access surveys' data from the HiPS2FITS service.[11] HiPS2FITS is a tool developed at CDS aiming at extracting FITS images from HiPS sky maps (P. Fernique et al. 2017; T. Boch et al. 2020), following the World Coordinate System (WCS) convention. This technique has several significant advantages in our use case, regarding dataset construction and image preprocessing. First, it allows us to generate image cutouts of the sky of any specified field of view, in a unified manner across surveys, regardless of the initial shot framing or pixel size. This accomplishes—and makes it much easier—photometric band alignment (image data of different photometric bands in VPHAS+ are not one-to-one), image tiling (to produce complete images of PNe), and some data augmentation (translation) all at once. Mosaicing, however, has a drawback when applied to input imaging data exhibiting background variations; it may introduce artifacts with rectangular patterns, corresponding to tile joints (Figures 9 and 10 illustrate the robustness of YOLO models to such artifacts). This problem was addressed on VPHAS+ by T. Boch et al. (2025) by applying careful background correction using the Montage toolkit. The technique they developed is applicable to other HiPS and could be profitably applied to IPHAS in the future.

We believe that our simple and unified data preprocessing pipeline is highly versatile and could be effectively applied to other machine learning tasks in astronomy, particularly given the extensive number of HiPS available online.

## 3. Methodology

Our aim is to detect PNe as light sources captured by IPHAS and VPHAS+ surveys. Our approach is to perform object detection on images built from cutouts of the sky surveys, within the supervised learning paradigm, using CNNs. This section describes in detail our methodology, including image preprocessing, image annotation, and deep learning techniques.

### *3.1. YOLO Deep Learning Model*

The YOLO approach in real-time object detection and classification was introduced in 2015 by J. Redmon et al. (2015). The model architecture, a CNN inspired by GoogLeNet, unified both bounding box and class probability predictions into a common regression task performed directly from full images in one evaluation by a single neural network. At that time, this simplified approach proved to be extremely fast and showed great performance, outperforming other state-of-the-art detection methods such as deformable parts models (DPMs) and region-based CNN (R-CNN) on various tasks. Since its first version, YOLO has continuously evolved (N. Jegham et al. 2025). Its latest version, YOLOv12, introduces a new attention-centric framework (see Y. Tian et al. 2025). We use the medium-size variant, YOLOv12 "m," comprising around 20.2 M parameters. It is pretrained on the Microsoft Common Objects in Context (COCO) dataset, a large-scale and widely used benchmark for object detection.

The model is designed to predict bounding boxes that fit the objects being searched for, as well as a class and a confidence score for each prediction.

### *3.2. Dataset Construction*

#### *3.2.1. Spectral Bands*

The YOLO model takes a three-channel image (RGB) as input—that is, as its first layer. We combine three spectral bands—$r$, $i$, and H$\alpha$—to produce multichannel images. The $r$ and H$\alpha$ bands both contain the H$\alpha$ spectral line, typical of PNe spectra (see Section 2.1). We do not use the $g$ band here, though it contains the O[III] spectral line, because it is not available in IPHAS. Exploring the use of $g$, for example as an alternative to $r$, may provide valuable insights in future investigations of VPHAS+. The $i$ band, in contrast, contains no major emission line of PNe. Its integration into the data is motivated by the idea that it should help the model to differentiate the signal (PNe) from other sources (stars, etc.) in the background. Figure 1 is an illustration of how efficiently this extra piece of information can be used to discriminate signal versus background, using a simple algebraic map.

[9] URL: http://www.iphas.org/
[10] URL: https://www.vphasplus.org/
[11] URL: https://aladin.cds.unistra.fr/hips/

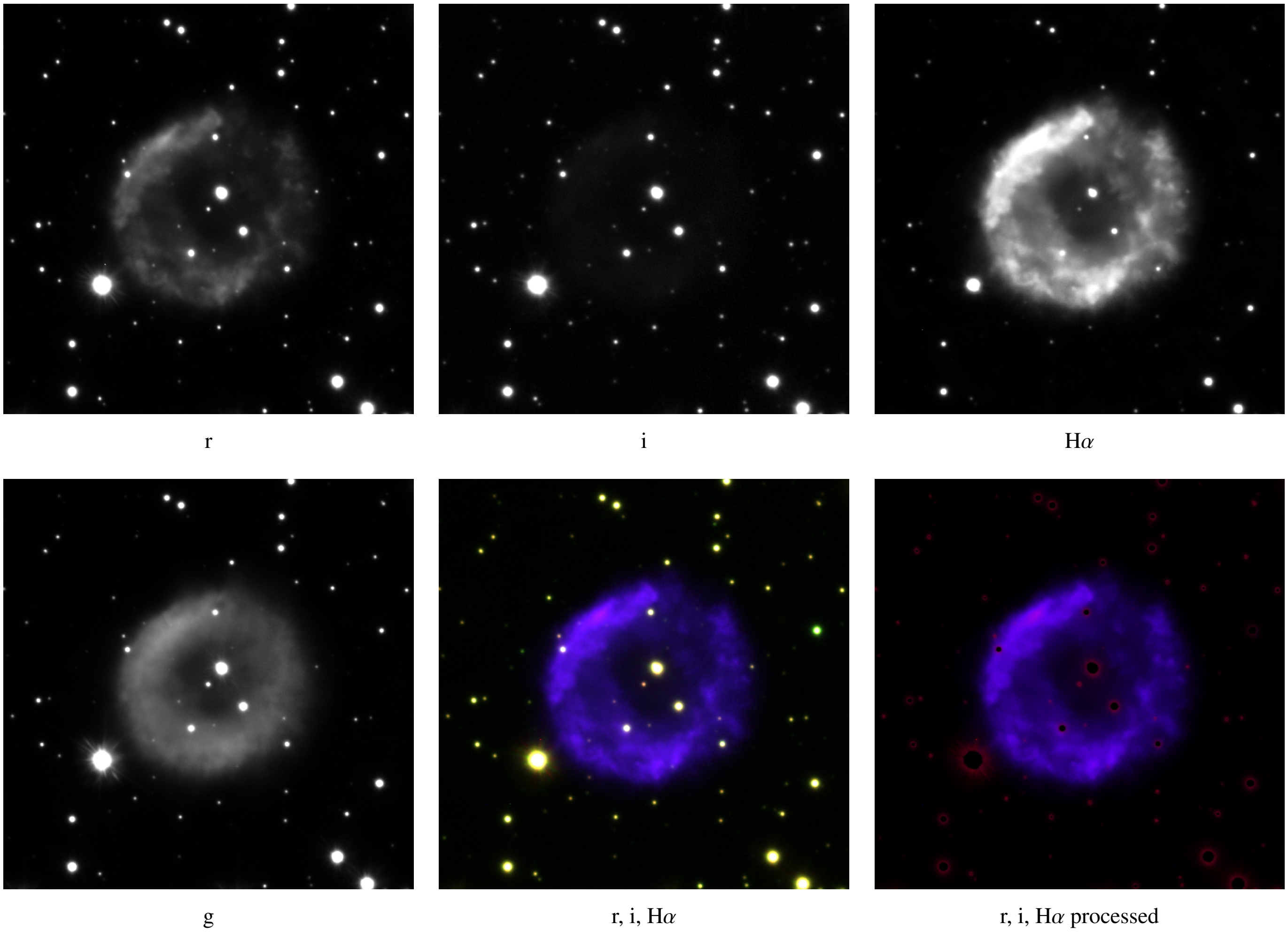

**Figure 1.** Small (640 × 640 px) images of NGC 2438 in various spectral bands (VPHAS+). The bottom right image is the r, i, H$\alpha$ image processed by applying the transformation $x \mapsto x \cdot (1 - i/255)$ to the $r$ and H$\alpha$ bands, and $x \mapsto 0$ to the $i$ band. This processing, which we do not use, is a mere illustration of how the $i$ band may be used to discriminate PNe signal versus background.

### 3.2.2. Field of View and Image Size

The model's computational complexity (and in turn, its training and prediction speed) considerably increases with the size of input images. On the other hand, PNe apparent diameters range from almost 0 to around 4.5 deg. Consequently, a trade-off must be made between the maximum size of targeted objects and the model speed. Using HiPS2FITS, we are able to generate cutouts of HiPS of any desired size and field of view, regardless of the CCD cameras' resolution and size in pixels. In order to take advantage of the full angular resolution of IPHAS and VPHAS+ surveys (see Table 1) while maintaining a reasonable model training speed, we worked with 640 × 640 px images (the default size in Ultralytics YOLO setting) with a field of view of ∼211 × 211 arcsec$^2$ ($3.5 \cdot 10^{-3}$ deg$^2$) and ∼134 × 134 arcsec$^2$ ($1.4 \cdot 10^{-3}$ deg$^2$), respectively. We decided to work in parallel with a twin dataset of 1280 × 1280 px images, with a field of view of 720 × 720 arcsec$^2$ (0.04 deg$^2$) and a degraded angular resolution equal to $0\farcs5625$ px$^{-1}$ in both surveys, in order to investigate more rare, extended objects as well. Those are potentially the most photogenic, which represents an additional interest for participants of the AIP-PN project.

Over 90% of cataloged Galactic PNe with diameter entered in HASH have apparent diameter below 134″. Nearly 99% are below 720″.

### 3.2.3. Image Preprocessing

We first generated HiPS cutouts of the desired sizes and fields of view, in the FITS format, from the CDS servers using HiPS2FITS (in the gnomonic projection). To each cutout in each spectral band, we applied histogram stretching and clipping before stacking them into three-channel 8 bit PNG images: $r$ (red channel), $i$ (green), and H$\alpha$ (blue). The stretching consists of performing a linear interpolation, mapping the first and the $(q - 1)$th of $q$ quantiles to 0 and 1, respectively, for $q = 200$. Then, we clip extreme values. Thus, the image is normalized between 0 and 1, and its lowest 0.5% and highest 0.5% values are set to 0 and 1, respectively. Finally, we normalize and discretize the image to an 8 bit encoding. The main purpose of stretching and clipping is to spread the signal over bits before applying conversion from float32 to 8 bit encoding. Indeed, since the signal is concentrated in the low intensity values of original images, it would be largely lost in the conversion process.

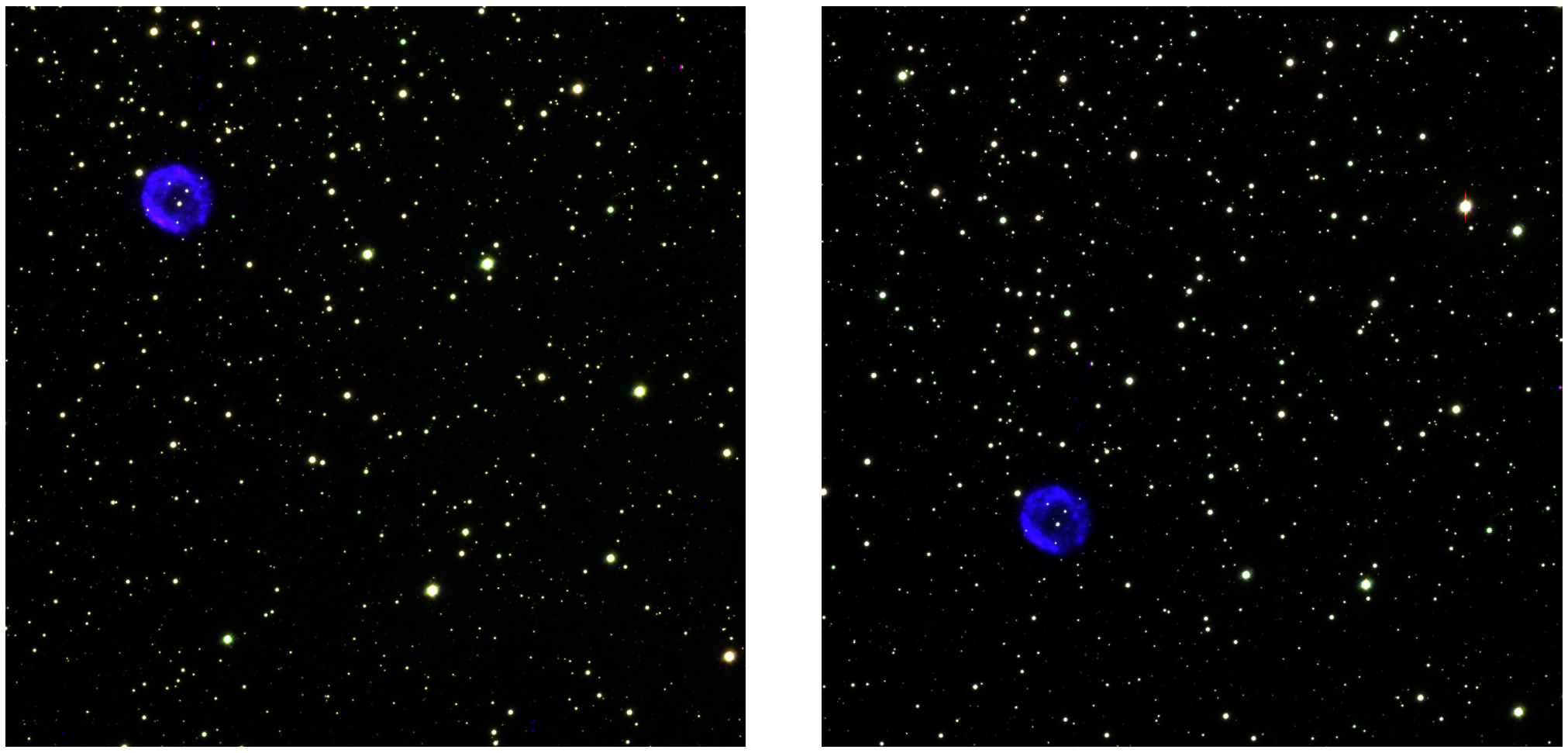

**Figure 2.** Large decentered images of NGC 2438 (1280 × 1280 px, VPHAS+).

Note that this process also improves the visual aspect of images. The latter could be further improved by replacing quantiles with other values adapted to each image individually, through IRAF's Zscale algorithm, for example. Though this consideration is relevant when it comes to visual inspection, it is not necessary in the deep learning framework because the model extracts features by itself.

#### *3.2.4. The Set of PNe*

The training, validation, and test sets are made up of images of PNe cataloged in the HASH database. From HASH, we select entries corresponding to Galactic PNe with status either True, Likely, or Possible, for which the major apparent diameter value is available and lower than the chosen image field of view width. Of course, we filter out PNe lying outside of the coverage of the surveys. Finally, the total number of selected PNe is 2596 for small images (564 are visible in IPHAS and 2103 in VPHAS+) and 2704 for large images (597 and 2180, respectively). Note that a number of them are visible in both surveys.

An image targeting each PN of the selection is generated. To avoid bias, we choose a random framing of the image so that the center of the targeted PN can occupy any position on the image, uniformly (note that a PN can end up partially out of frame; see Figure 10). Experimentation has shown that this decentering is essential for the model to detect PNe anywhere in the images. Note that the field of view of an image sometimes contains several cataloged PNe, due to their proximity in the sky.

We also produce a second randomly framed image of each NP to the dataset for data augmentation purposes (see Figure 2). Thus, the full datasets of small and large images comprise respectively 5334 and 5554 images, from which we extract 60% slices for model training.

#### *3.2.5.* k*-fold Cross Validation and Train–Validation–Test Splits*

We then perform a $k$-fold split for cross validation, with $k = 4$, and a train (60% of the selected PNe), validation (15%), and test (25%) split. These splits are performed *on the list of PNe unique identification numbers* (the "idPNMain" feature in HASH), rather than on image files. This is important to avoid data leakage from the train set to the test set, since a PN sometimes appears on both surveys (this is sometimes called a grouped $k$-fold split) and appears on two randomly framed images. We stratify the $k$-fold split according to the PN status, so that instances of these statuses are distributed evenly among the folds.

#### *3.2.6. Image Annotation*

Instead of manually labeling each image of the dataset, we use a more efficient automated procedure relying on the HASH database. Whenever an image contains the position of a cataloged PN, we create a square bounding box centered at this position, with side length equal to the apparent diameter (the database's "MajDiam" feature) of that PN, each side being parallel to a border of the image. A text file is created alongside that image, containing the coordinates of all corresponding bounding boxes in the YOLO "xywh" format. Those are called "ground-truth" labels, as opposed to the labels predicted by the model.

This baseline procedure has several limitations and shortcomings, though. First, the apparent diameter value is not always provided in the HASH database. Second, the image depicting a cataloged PN may very well contain an uncataloged PN in its field of view; the labeling is thus (probably marginally) incomplete. Third, the bounding box is often poorly adjusted to the actual silhouette of the associated PN. Indeed, PNe assume a variety of very different morphologies, elliptical, round, bipolar, etc., and we are drawing only squared bounding boxes. In addition, the labeling was not made directly on the images of the surveys, though the apparent diameter depends on the photometric band in which the observation was made, the sensitivity of the instrument used, optical artifacts around bright sources, etc. Some PNe cataloged in HASH are in fact too faint to be visible on the surveys we use, and image quality varies significantly throughout surveys.

**Table 2**
Ultralytics YOLO Data Augmentation Parameters Used for Training

| Parameter | Value | Description |
|---|---|---|
| `hsv_s` | 0.2 | Maximum changes to saturation (fraction) |
| `hsv_v` | 0.2 | Maximum changes to value (fraction) |
| `translate` | 0.1 | Maximum translation (fraction of image size) |
| `scale` | 0.2 | Maximum image scaling (zoom in/out) |
| `flipud` | 0.5 | Vertical flip probability |
| `fliplr` | 0.5 | Horizontal flip probability |
| `degrees` | / | Customized rotations: $-180°$, $-90°$, $0°$, or $90°$ |

It should be kept in mind that the blurred nature of PNe makes their annotation by humans quite subjective, and biased by image processing techniques affecting contrast, such as IRAF's Zscale linear histogram stretching algorithm.

To tackle these difficulties, we adapted YOLO models' loss weights, as well as our evaluation methodology, to insist on the importance of object versus background detection rather than on accurate prediction of the shape of each bounding box (see Section 3.3 and Section 4.1). See also Section 4.3 for further discussion on this point.

#### 3.2.7. Data Augmentation

A first data augmentation consists of the generation of a second randomly framed image of each PN, as discussed in Section 3.2.4. More than 70 selected PNe are visible in both surveys, which also represent a data augmentation. We use the built-in functions of Ultralytics YOLO for further data augmentation (see Table 2). We customize the Ultralytics implementation in order to apply random rotations, chosen uniformly in $\{-180°, -90°, 0°, 90°\}$, to the images.

### 3.3. Model Performance Evaluation Metrics

Let us briefly recall the standard evaluation metrics used in object detection tasks, and in particular by YOLO models.

The intersection over union (IoU) of two rectangles $A$ and $B$ (e.g., two bounding boxes) is defined as the ratio $\mathrm{IoU}(A, B) = \frac{\mathrm{Area}(A \cap B)}{\mathrm{Area}(A \cup B)}$. A predicted bounding box will be considered either a true positive (TP) or a false positive (FP) depending on the value of its IoU with ground-truth bounding boxes: greater than or less than a given IoU "matching" threshold, respectively. A ground-truth box having lower IoU with all predicted boxes is considered a false negative (FN). In this context, images with no ground-truth box and no detection at all are commonly considered as a true negative (TN). IoU is also useful to avoid multiple predictions of the same object by applying nonmaximum suppression (NMS), removing overlapping boxes (with IoU over an "NMS" threshold) by retaining the box with the highest confidence score.

As discussed in Section 3.2.6, the definition of a bounding box for a PN image is not always clear; our labeling strategy is simple but imperfect. Therefore, we should not expect the model to reproduce the exact shape of these boxes; what is important to us is the detection of PNe on images, with a rough indication of their location rather than their precise shape. We set both IoU thresholds to 0.01 and perform NMS only after ground-truth matching.

The confusion matrix is, in fact, a mere table gathering the TP, FP, TN, and FN counts in the prediction results of a particular model over a given dataset. It depends on the IoU thresholds and confidence threshold. We count TP without multiplicity; that is, two TPs targeting the same ground-truth box are counted only once.

The precision $P = \mathrm{TP}/(\mathrm{TP} + \mathrm{FP})$ accounts for the proportion of positives that are true, while the recall $R = \mathrm{TP}/(\mathrm{TP} + \mathrm{FN})$ accounts for the proportion of true (meaning true objects) that are positives. The F1-score is the harmonic mean of the precision and recall.

Both precision and recall depend on the confidence threshold $C$, and the image of the map $p$: $[0, 1] \to [0, 1]^2$, $C \mapsto (R, P)$ may in practice be interpreted as a subset of the graph of the precision as a function of the recall. This is called the precision–recall (PR) curve. The integral of this function is a very synthetic evaluation metric, called the average precision (AP; interpreted as the mean of the precision function because the recall range is an interval of length 1), which does not depend on any confidence threshold but still depends on both the IoU "matching" and "NMS" thresholds.

## 4. Results

### 4.1. Models Training and Hyperparameters Tuning

Our models were trained on a 40 GB NVIDIA A100 graphics processing unit (GPU) for 200 epochs on small (resp. large) images, within around 3.5 (resp. 24) hours, using the AdamW optimization algorithm (with batch sizes equal to 20 and 3, resp.). Model hyperparameters were tracked using MLflow during the tuning process on small images. The best hyperparameters found were then used on large images as well, for which a full tuning process would be too costly. Table 2 presents the augmentation parameters we used.

YOLOv12's loss function is defined as a linear combination of three losses: the box loss, the class loss, and the distributed focal loss (DFL; mostly beneficial in imbalanced multiclass situations). Because we are working with a single class of objects (PNe), and because we place more importance on object versus background detection than on accurate prediction of the shape of each bounding box (see Section 3.3), we set the loss weights to 1, 8.25, and 0.25, respectively. Indeed, with a single-class dataset, the class loss component should confirm the presence of the class versus a negative (background) example.

The graphs in Figure 3 show that our model converges after 160 (resp. 130) epochs on the small (resp. large) images dataset, indicating that training can be stopped at these steps. When examined separately, however, box loss and class loss seem to indicate overfitting from the hundredth epoch onward. The fact that the AP (computed on the validation set with IoU thresholds of 0.01) keeps improving afterward confirms the importance of class loss over box loss and DFL loss in our context. Variations in the graph corresponding to large images might be explained by the low batch size used.

### 4.2. Models Performances

Though our model was trained on images depicting PNe of arbitrary status (True, Likely, and Possible), we evaluate it on True PNe images only, in particular to allow comparison with the existing literature. We evaluate the model on a dataset without augmentation (except that it contains images from both overlapping IPHAS and VPHAS+ surveys). Considering the PR curves, we set the confidence thresholds to 0.15 for confusion matrix computation. Table 3 presents the results of the $k$-fold cross-validation procedure achieved on the small image dataset.

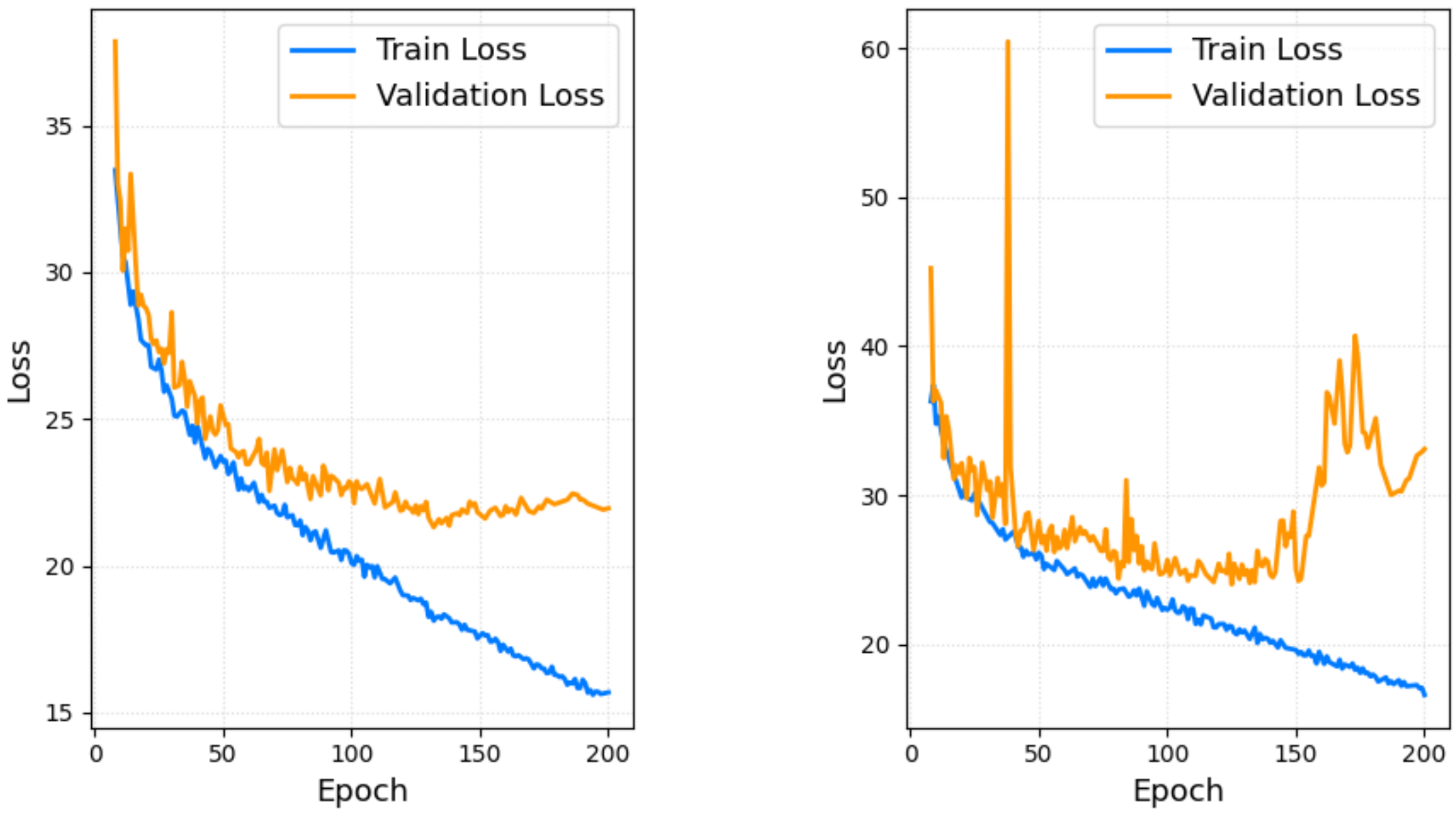

(a) For small images

(b) For large images

**Figure 3.** Loss function graphs for small and large image datasets. These graphs were obtained during the training of YOLOv12 on the first cross-validation folds.

We present further details concerning the first cross-validation fold of each dataset in Figures 4 and 5.

The APs at IoU threshold 0.01 (AP01) are 0.979 and 0.950 for small and large image datasets, respectively. The model performs better on small images than on larger ones. This difference may be explained by variations in angular resolution, which can affect the recall rate (by making small-object detection more challenging), and variations in field of view, which can affect the precision rate. Indeed, large images have a field of view approximately 12 and 29 times larger than small IPHAS and VPHAS+ images, respectively, and thus contain a much larger number of celestial sources (hence a smaller ratio of PNe). In addition, with large images we are targeting more extended, hence generally fainter, PNe.

A total of 506 small images and 516 large images were analyzed by our models, with a processing time of around 18 and 200 ms per image, respectively. The performances are extremely good (see Figure 5); on small images, our model achieved a precision rate of 99.0% and a recall rate of 97.3%. The performance on large images is 94.8% (precision) and 94.8% (recall).

**Table 3**
YOLOv12m Model $k$-fold Cross-validation Results ($k$ = 4) on the Small Image Dataset

| Fold | AP01 | Precision | Recall | F1-score |
|---|---|---|---|---|
| #1 | 0.979 | 99.0% | 97.3% | 98.1% |
| #2 | 0.996 | 99.6% | 98.4% | 99.0% |
| #3 | 0.990 | 98.0% | 97.4% | 97.7% |
| #4 | 0.974 | 98.0% | 96.4% | 97.2% |
| **Mean** | **0.985** | **98.7%** | **97.4%** | **98.0%** |
| Standard deviation | 0.010 | 0.790 | 0.818 | 0.761 |

**Note.** The confidence score threshold is set to 0.15.

### 4.3. Discussion

*Visual analysis of FP and FN.* A visual analysis of the five FPs found after training the model on the first cross-validation fold of the small image dataset (see Figure 5) shows that two of them are caused by CCD defects in IPHAS, having a morphology close to that of a compact PNe (compare Figures 6 and 9(a)), and one is in fact targeting a true PN (called "Th 4–3") with its apparent diameter value incorrectly set to 0 in the HASH database (thus leading to a zero IoU value). As a consequence, this object is also incorrectly considered as an FN. Among the 14 FNs, 2 are due to overlapping ground-truth bounding boxes, due to duplicates of PNe in the HASH database (indexed under the same "idPNMain" identifier).

*Small-object detection performance.* Among the 70 ground-truth boxes with a diagonal below 30 pixels in the first cross-validation fold of the small image dataset (see Figure 5), only 64 were detected by the model, yielding a recall rate of approximately 91.4%. In contrast, larger bounding boxes achieved a recall rate of approximately 98.2%, suggesting that the model exhibits reduced performance on smaller objects.

*Label quality and low IoU threshold choice.* Although the ground-truth bounding boxes used in this study were not produced by manual annotation, they are derived from the HASH PN catalog, which represents the most comprehensive and widely used compilation of PNe properties currently available. These measurements, gathered from multiple observational programs and literature sources, inevitably reflect a diversity of methodologies and definitions. While this heterogeneity introduces a degree of morphological uncertainty, it also captures the realistic variability encountered in large-scale survey data and ensures that the detector is trained and evaluated under conditions representative of real-world applications. In particular, the use of catalog-based diameters means that bounding boxes may encompass faint halos or extended structures that are only partially visible in IPHAS or VPHAS+ imagery. For this reason, localization metrics must be interpreted with an understanding of this intrinsic uncertainty, and the adoption of a low IoU threshold is a deliberate methodological choice to prioritize robust detection of PNe presence over geometric precision. Although PNe apparent diameters are not always precisely defined, we place our trust in the HASH database regarding the position of their centers. Figure 7 shows that a significant number of TPs are indeed detected by our model with an IoU value below 0.5 (see also the examples of detections in Figures 9 and 10). Far from undermining the validity of the results, this approach aligns the evaluation criterion with the scientific objectives of PNe searches, where reliable candidate identification within a field of view is more critical than pixel-level boundary accuracy, and provides a baseline for future improvements, such as the development of manually curated subsets for precision benchmarking. These improvements could build on an automated procedure for computing bounding boxes around deep-sky objects in astronomical images such as the algorithm proposed by O. Parisot & M. Jaziri (2024). However, the objective of the present work is not to propose a method for completing the HASH database by performing accurate image segmentation of PNe, refining the calculation of their apparent diameters, or computing other astrophysical characteristics (e.g., morphological classification). The great morphological variability of PNe limits the use of physical models and leads us to adopt an astronomical observation perspective instead; this is why we employ YOLO models. Through their statistical approach, we can aim for PN detection without predefined morphological criteria. This approach is aligned with our goal of detecting PNe to enable new discoveries, which will lead to a new observation campaign.

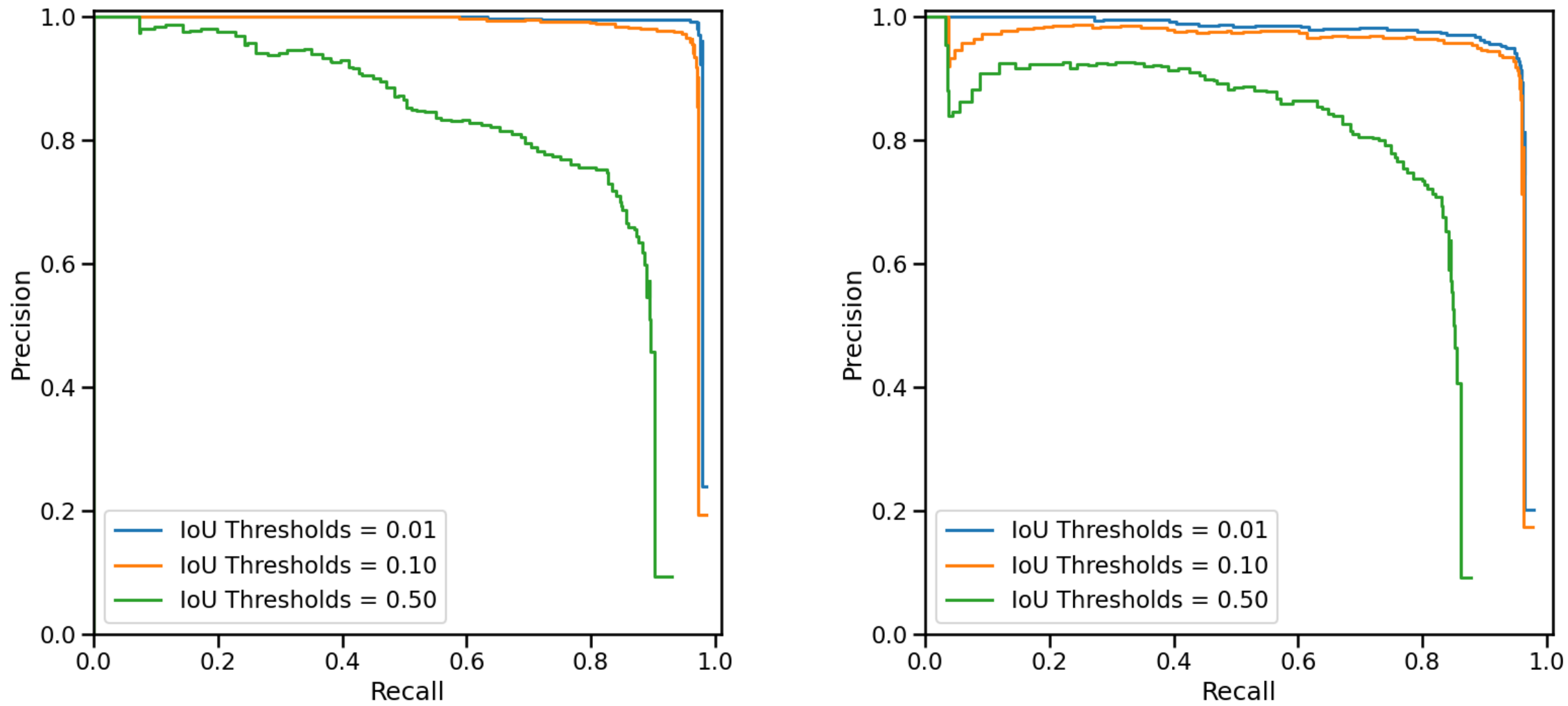

(a) For small images

(b) For large images

**Figure 4.** PR curves for small and large image datasets. The confidence score threshold step is 0.01. For small images, AP01 = 0.979, AP10 = 0.970, and AP50 = 0.792. For large images, AP01 = 0.950, AP10 = 0.934, and AP50 = 0.745.

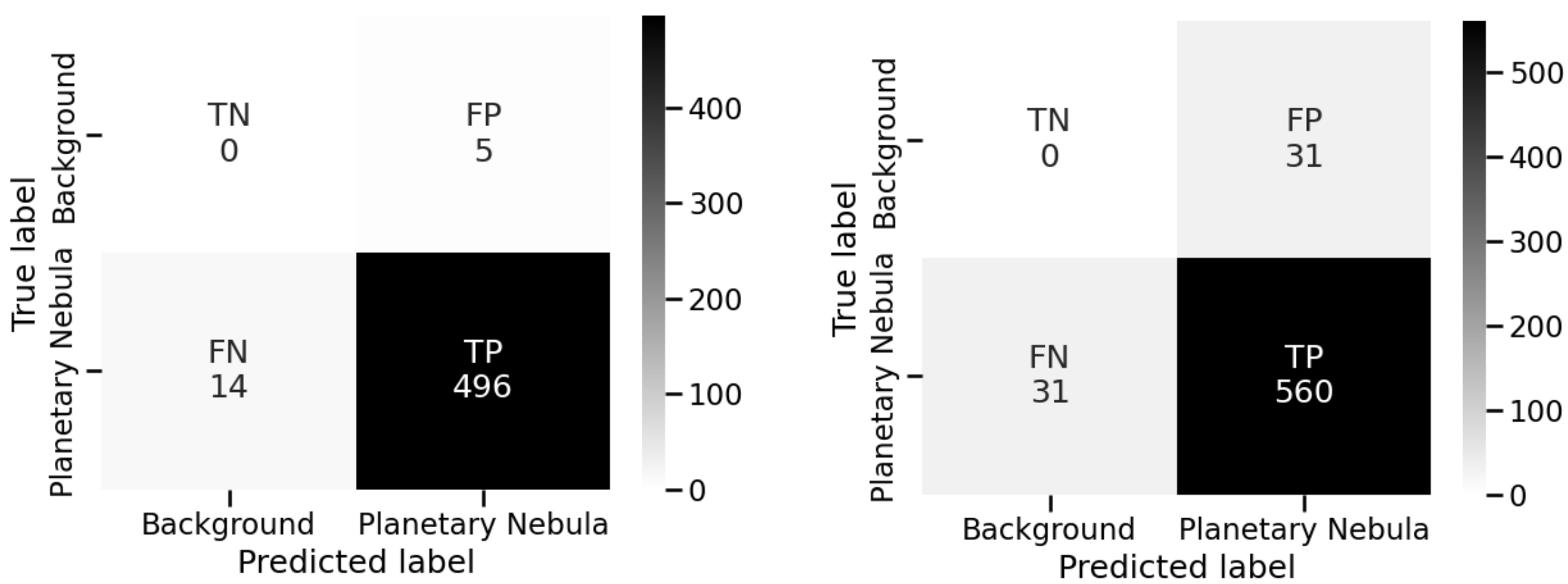

(a) For small images

(b) For large images

**Figure 5.** Confusion matrices for small and large image datasets.

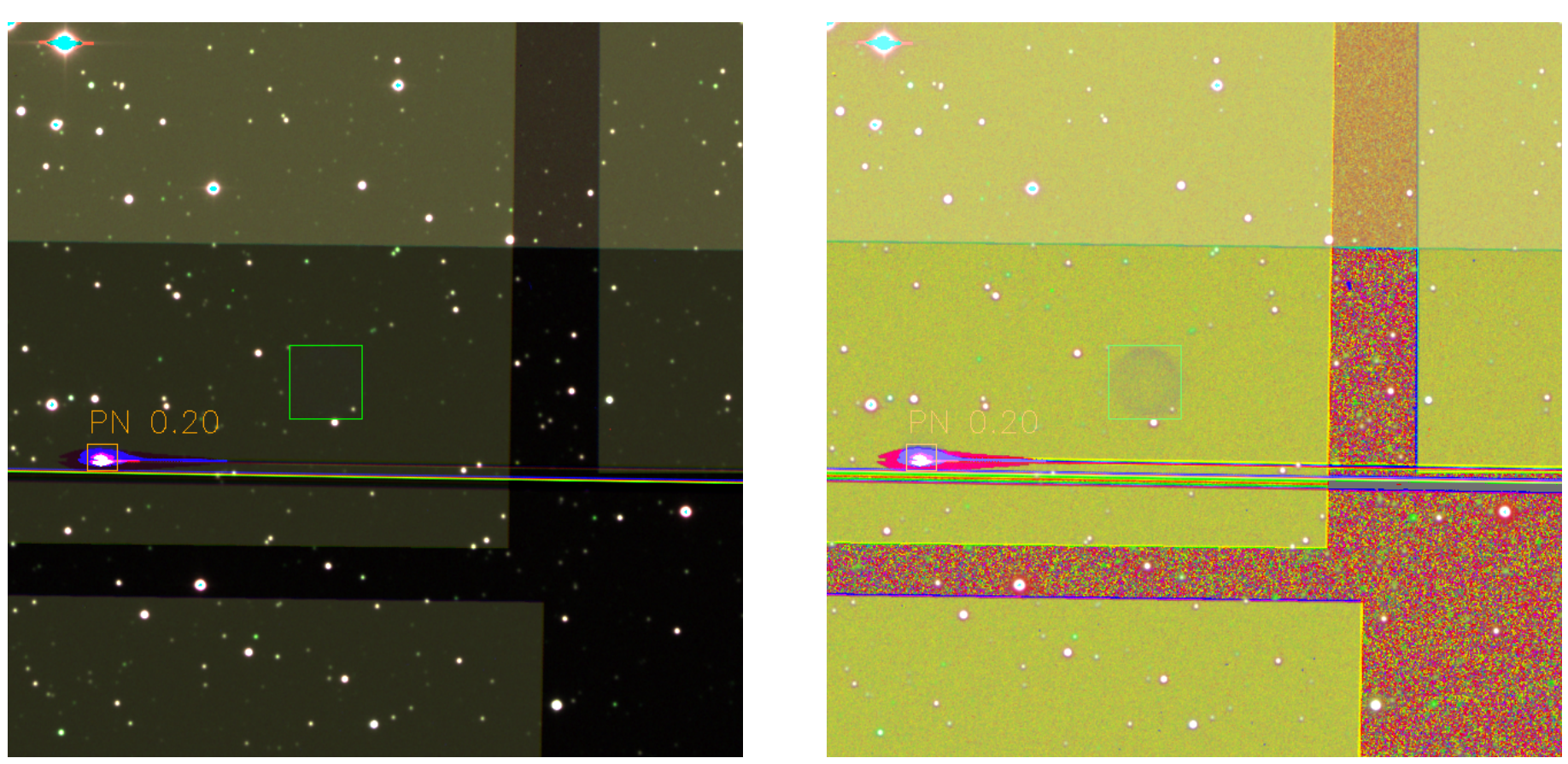

**Figure 6.** Example of an FP with confidence score 0.20 (orange bounding box) due to a CCD defect in IPHAS. Pushing the contrast and brightness of the image highlights the presence of an FN (PN "LDu 18") in the central part of the image (green bounding box).

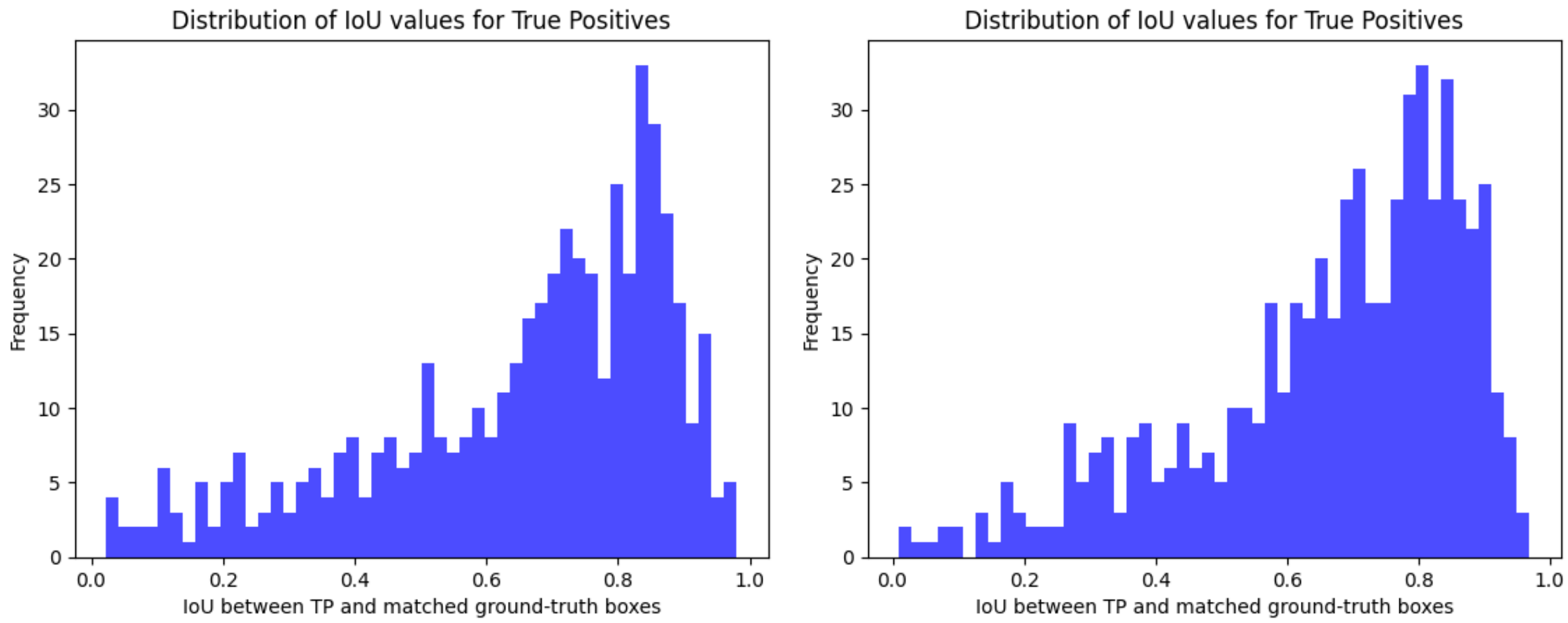

(a) For small images

(b) For large images

**Figure 7.** Distribution of IoU values for TPs and their matched ground-truth bounding boxes for small and large image test datasets (concerning the first cross-validation fold of each dataset).

In most conventional object detection tasks, model performance is typically evaluated at relatively high IoU thresholds, often 0.5 or greater, reflecting the importance of precise spatial localization. However, PNe pose a fundamentally different type of detection problem. They are extended, diffuse, and frequently asymmetric objects whose visible boundaries are often ill-defined, particularly near the detection limit or in crowded Galactic fields. Adopting a deliberately lower IoU threshold (0.01) therefore constitutes a scientifically motivated choice tailored to the characteristics of the problem. This metric emphasizes the model's ability to identify PN candidates even when their morphology is irregular or their precise spatial boundaries are uncertain. It also aligns closely with the practical objectives of large-scale PN searches, where the primary goal is to build robust candidate catalogs for subsequent human verification and spectroscopic follow-up. Framing the evaluation in this way ensures that the reported performance reflects the true operational value of the detector and captures the aspects of detection that are most relevant for advancing PN science.

*Artefacts and PNe mimics impact on model performance.* The presence of astronomical and instrumental structures that mimic the appearance of PNe is an inherent and well-recognized aspect of optical narrowband surveys. Compact H II regions and supernova remnants, for example, can exhibit morphologies and spectral characteristics that closely resemble those of PNe, while instrumental effects such as CCD defects, ghost images, tiling edges, or variable background gradients may introduce additional complexity. This diversity of potential FPs highlights the difficulty of the detection task and underscores the importance of designing systems that can operate effectively under realistic survey conditions. The detector presented here achieves robust performance on confirmed PN detections despite this challenging environment, demonstrating that it captures meaningful morphological and photometric features even in the presence of strong confounding factors. Figure 8 indicates that detections with high confidence scores are predominantly TPs, whereas detections with low confidence scores comprise a mixture of TPs and FPs. This pattern highlights the challenging trade-off between precision and recall when selecting an appropriate confidence threshold. It is also worth noting that the set of FPs may include potentially interesting PN candidates.

As previously noted in Section 4.2, the precision metric depends heavily on the ratio of PNe to other sources in the images. Each test image contains at least one PN, which does not reflect the actual distribution of PNe across the galaxy. Furthermore, this ratio is sensitive to the field of view covered by each image, as it affects the extent of the background part. This is why it is crucial to deploy the model on the largest scale possible to obtain a meaningful measurement of its precision. We deployed the model to the full IPHAS and VPHAS+ surveys, as described in Section 4.4. The results of this large-scale deployment are discussed in the conclusion.

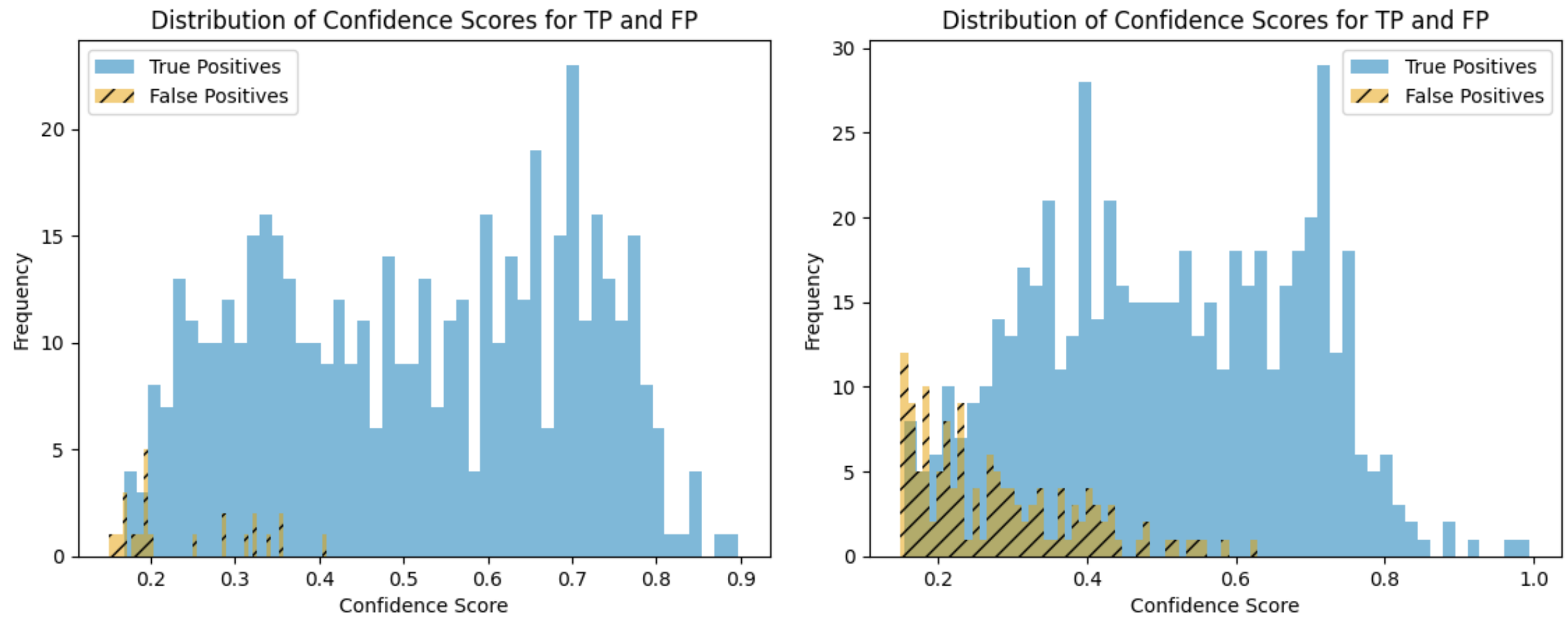

(a) For small images

(b) For large images

**Figure 8.** Distribution of confidence scores for TP and FP for small and large image test datasets (concerning the first cross-validation fold of each dataset).

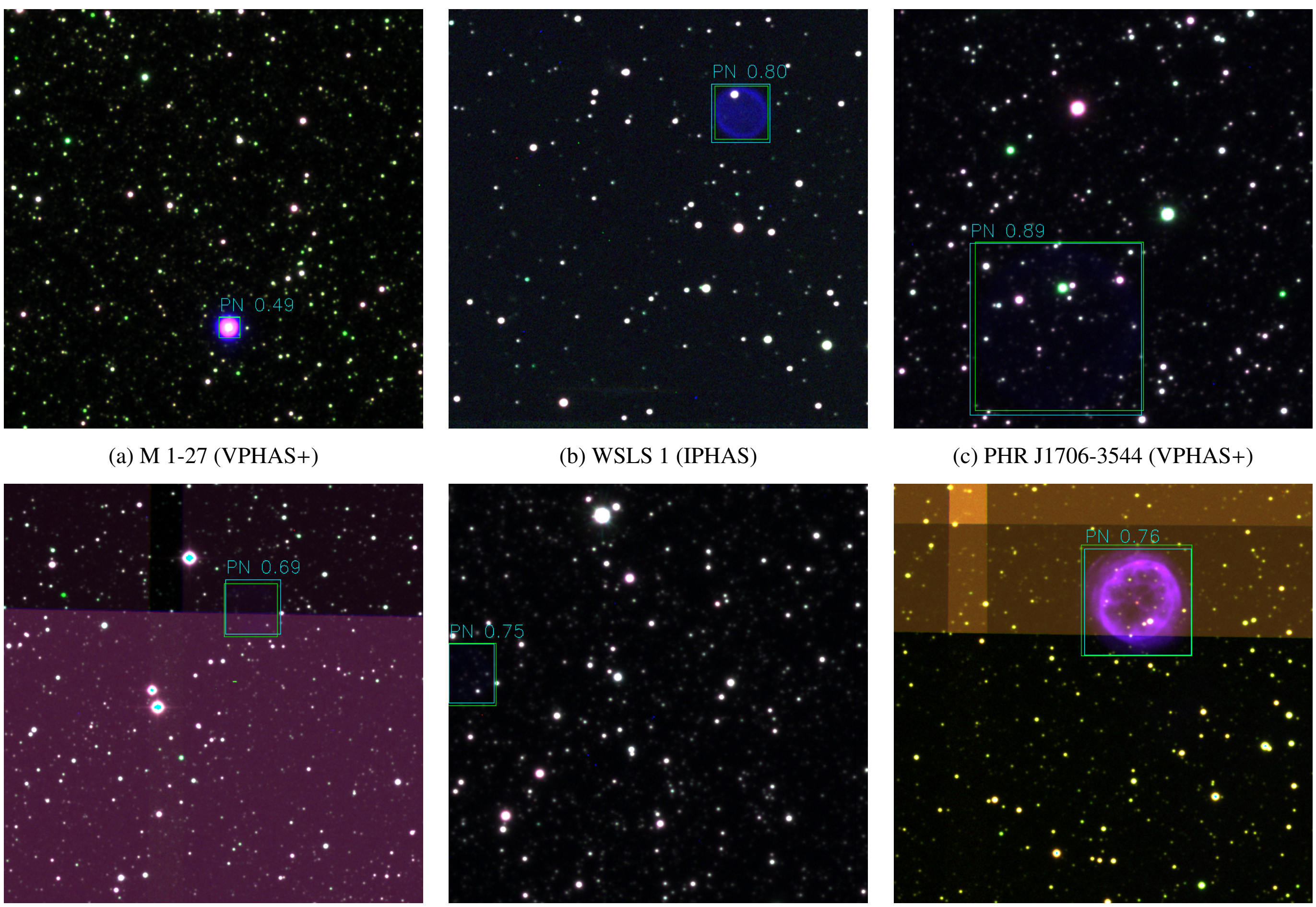

(a) M 1-27 (VPHAS+)

(b) WSLS 1 (IPHAS)

(c) PHR J1706-3544 (VPHAS+)

(d) IPHASX J193308.9+155354

(e) PHR J1734-2000 (VPHAS+)

(f) NGC 6894 (IPHAS)

**Figure 9.** Small images selected from the test set.

*Examples of detections (TP).* Illustrative visualizations of the detection results are shown in Figures 9 and 10. Figure 9 shows that our detection model accurately identifies PNe (green bounding boxes) and determines their positions (blue bounding boxes, including confidence scores). Figure 10 shows illustrative examples of the robustness of our detection model to artifacts such as CCD corner defects and HiPS tiling, particularly in IPHAS cutouts. It also highlights the importance of using a low IoU threshold, given the diffuse and cloudy morphology of PNe.

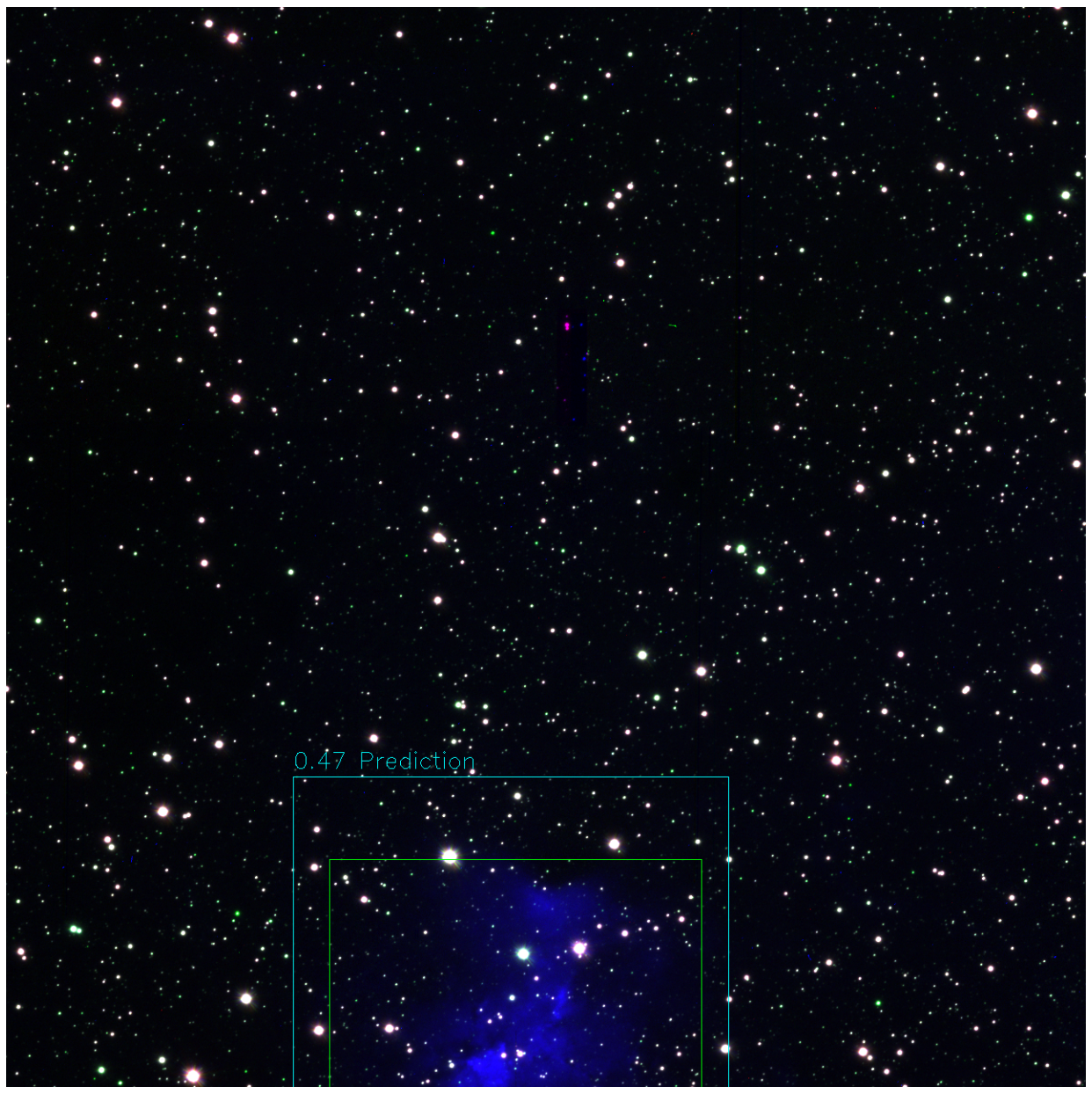

(a) RCW 69 (VPHAS+)

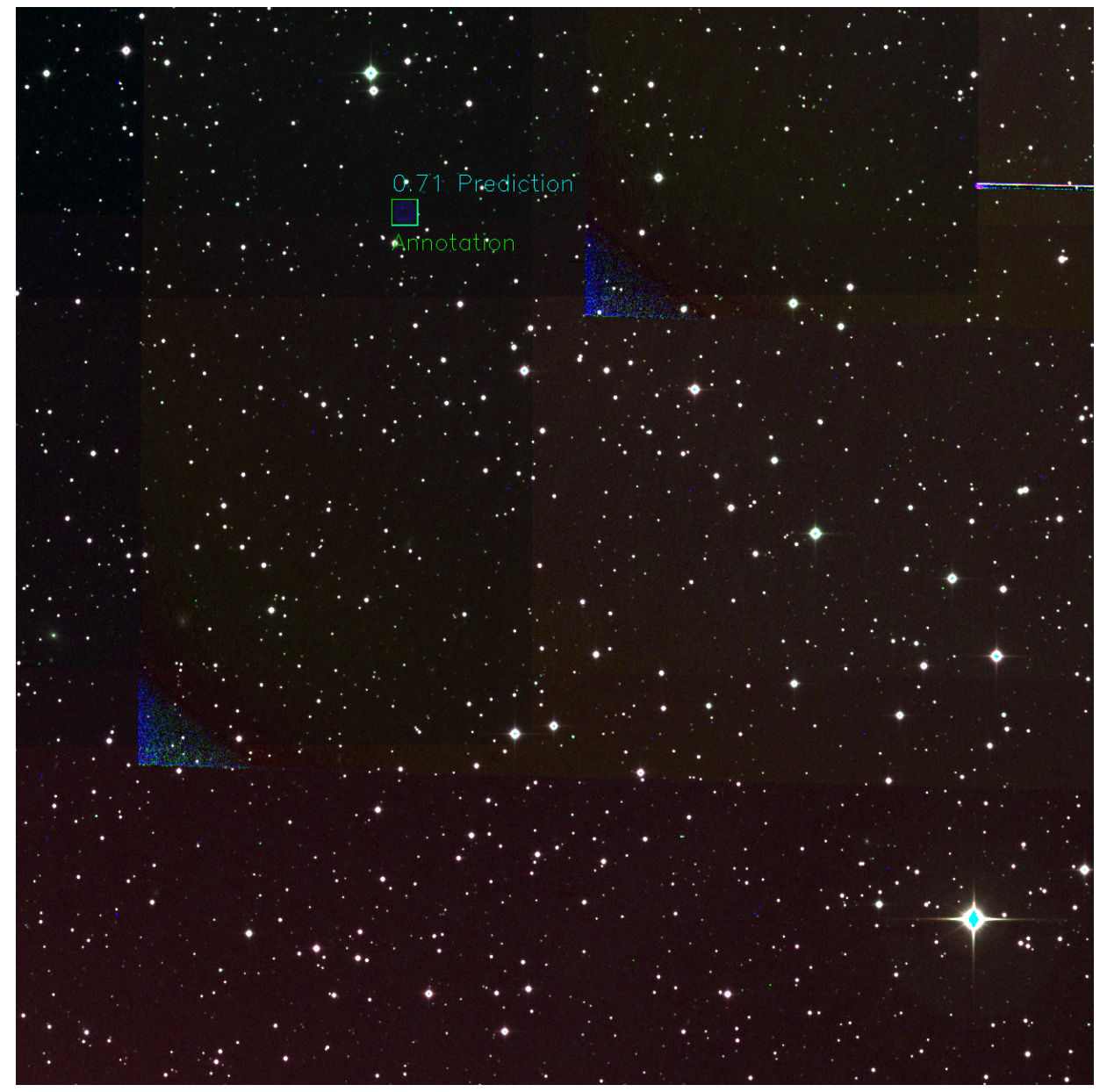

(b) IPHASX J045847.7+373640 (IPHAS)

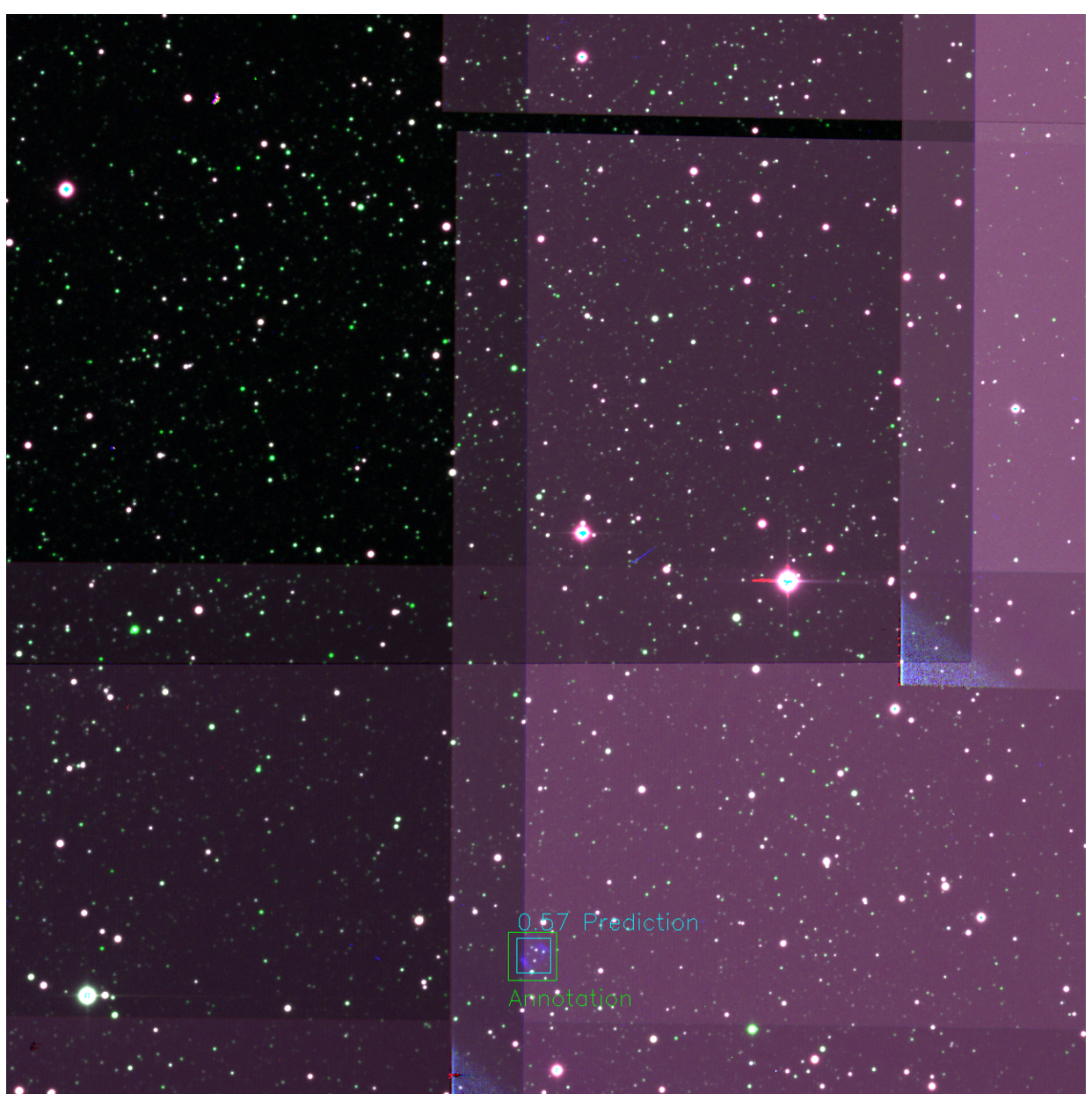

(c) IRAS 19086+0603 (IPHAS)

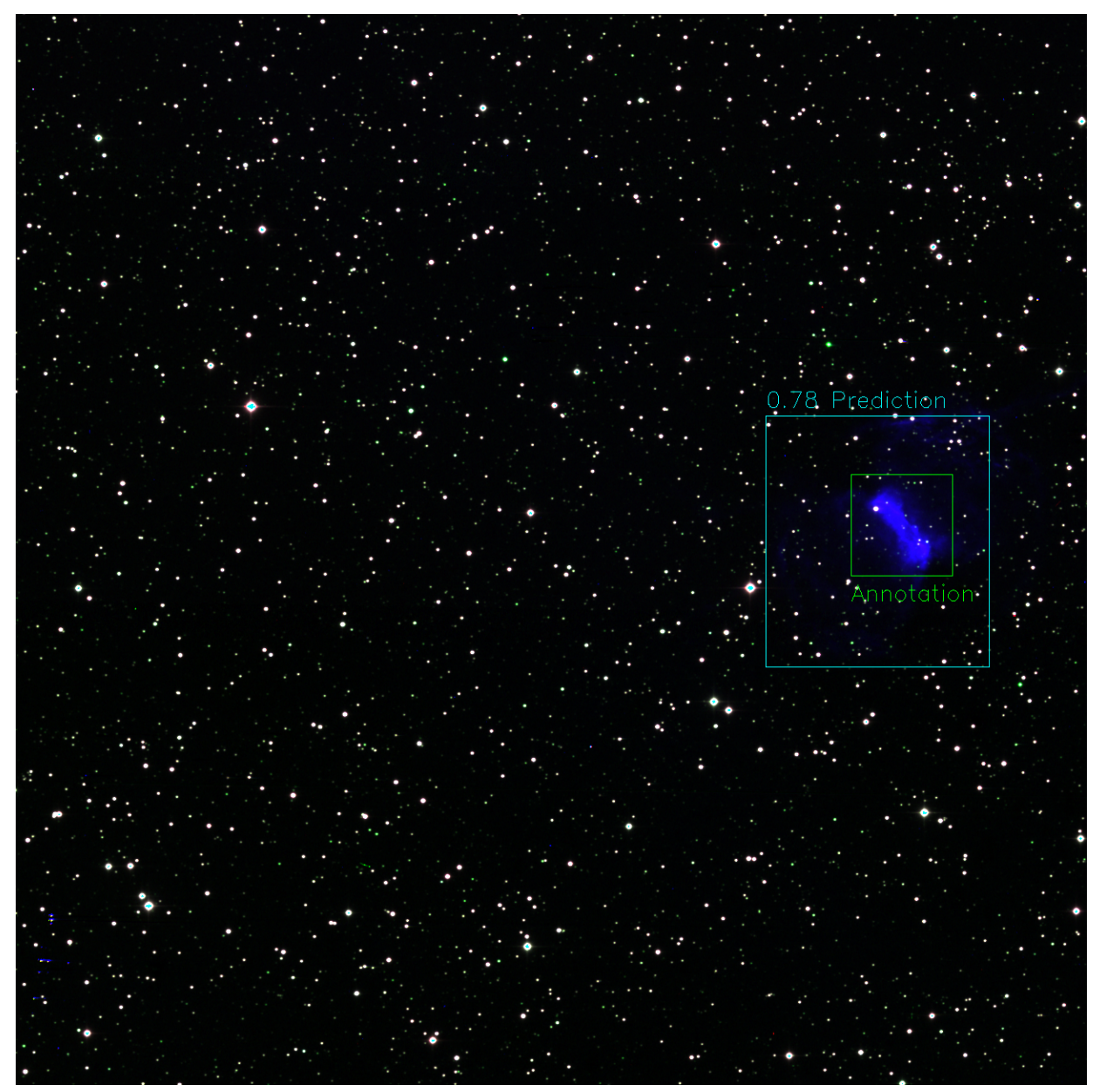

(d) Sh 1-89 (IPHAS)

**Figure 10.** Large images selected from the test set.

*Explainability with Eigen-CAM.* In addition to evaluating the model's quantitative performance, Eigen-CAM (M. Bany Muhammad & M. Yeasin 2021) was employed to gain further insight into the internal decision-making process of the YOLOv12 deep learning model. This technique provides a visual explanation by highlighting the most influential regions in the input that contribute to the model's predictions. Examining these activation maps shows that the model's detection head focuses on meaningful PN features across scales: envelope, boundary, background, etc.

YOLOv12's architecture includes multiscale detection heads that specialize in detecting objects of various sizes, allowing the model to effectively localize compact and extended PNe as illustrated by the activation maps of layers 4, 6, and 19 in Figure 11. We used Shirvalkar's implementation,[12] which is based on Gildenblat's package (J. Gildenblat et al. 2021).

*Comparison of performance between IPHAS and VPHAS+.* Table 4 shows that our model, trained on both surveys, performs better on VPHAS+ than on IPHAS. This discrepancy might be explained by lower image quality in IPHAS and background correction in VPHAS+ (see Section 2.4).

*Comparison with the literature.* In their work, R. Sun et al. (2024) used 512 × 512 px cutouts (with a field of view equal to 170 × 170 $\mathrm{arcsec}^2$) of the original 2 048 × 4 096 px IPHAS images to train and test their Swin Transformer model, before

[12] URL: https://github.com/rigvedrs/YOLO-V12-CAM

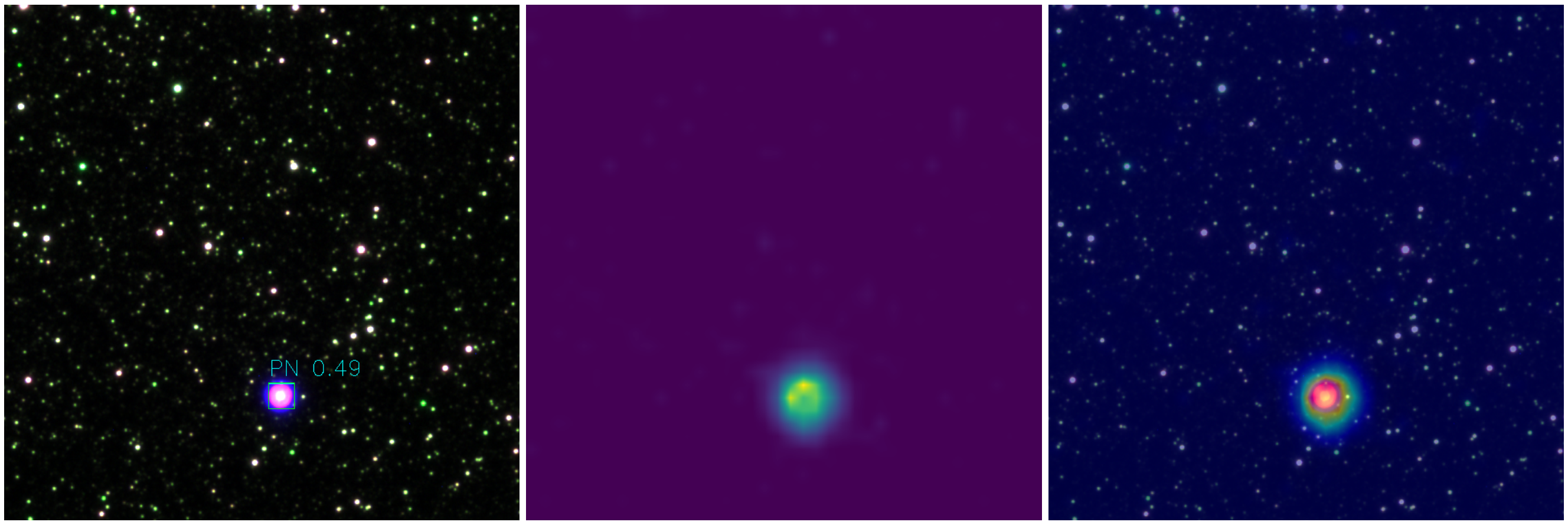

(a) M 1-27 (VPHAS+), YOLOv12m's layer 4.

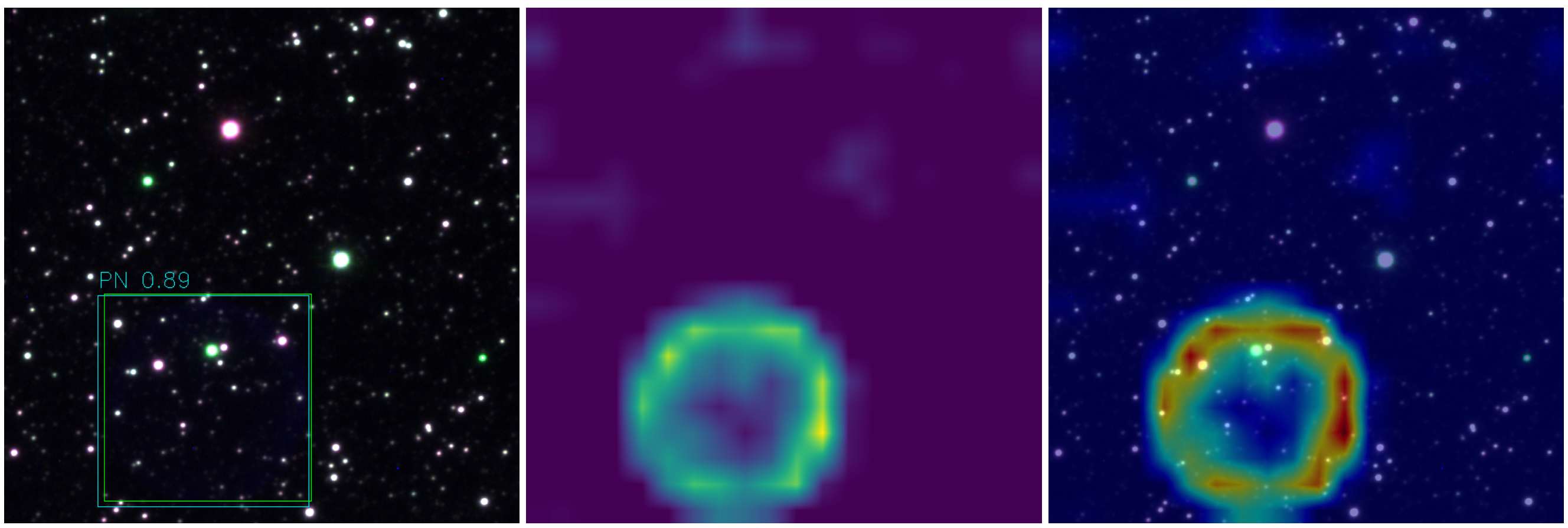

(b) PHR J1706-3544 (VPHAS+), YOLOv12m's layer 6.

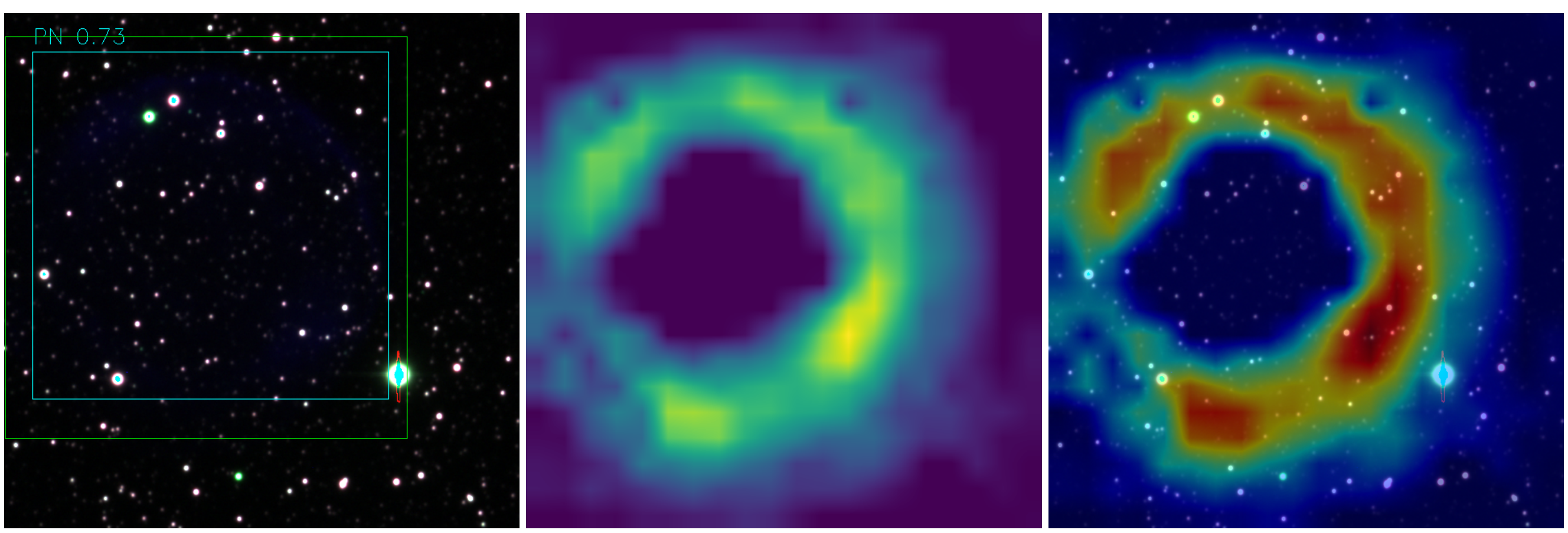

(c) Abell 62 (IPHAS), YOLOv12m's layer 19.

**Figure 11.** Original 640 $\times$ 640 px image (left), Eigen-CAM activation map (center), and their superposition (right).

applying it to the VPHAS+ survey. They preprocessed images using the Zscale algorithm in order to accelerate model convergence, which was achieved after around 40 epochs. They combined the $r$ and H$\alpha$ spectral bands into RGB images (with RGB = rrH$\alpha$). After training their model on a set of 1137 images, they tested it on a set of 454 images, both sets of which contain known True PNe. Their model achieved an AP of 0.944 when the IoU threshold is set to 0.5. Using an unspecified confidence score, they obtained a precision rate of 96.5% and a recall rate of 97.8% (444 TP, 10 FN, and 16 FP). As previously noted, when comparing results, it should be kept in mind that the field of view of our images is much larger than that of theirs.

*Comparison with YOLOv11.* Table 5 shows the results obtained with YOLOv11, allowing comparison with YOLOv12 (see Table 3). Results are almost identical for both versions of the model, suggesting that the latest modifications in the YOLO architecture, in particular the introduction of an attention

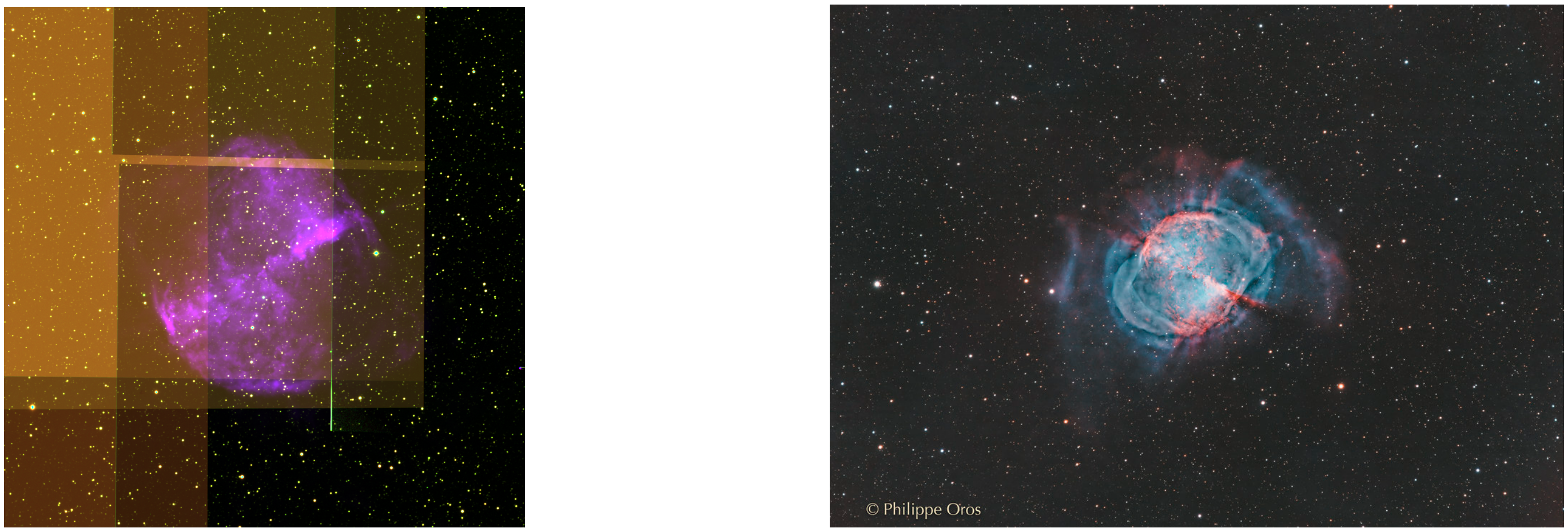

**Figure 12.** On the left, Messier 27 pictured by IPHAS (with HiPS tiling). Standard exposure times are 30 s ($r$), 10 s ($i$), and 120 s (H$\alpha$). On the right, Messier 27 pictured by P. Oros: Data were acquired with a CFF 200 mm f/6.5 apochromatic refractor equipped with a Televue 2x amplifier and a QHY600 M-PH monochrome camera. The dataset consists of 96 × 300 s narrowband frames (H$\alpha$ and O[III], combined in HOO) and 116 × 120 s broadband frames (R, G, B), for a total integration time of 11 hr 52 minutes. Observations were conducted from a backyard site in Mougins, France.

**Table 4**
YOLOv12 Model $k$-fold Cross-validation ($k$ = 4) Results on IPHAS and VPHAS+ Images of the Small Image Dataset

| | AP01 | Precision | Recall | F1-score |
|---|---|---|---|---|
| **Mean (IPHAS)** | **0.977** | **97.7%** | **95.9%** | **96.8%** |
| Standard deviation | 0.019 | 1.90 | 1.79 | 1.53 |
| **Mean (VPHAS+)** | **0.986** | **98.9%** | **97.8%** | **98.3%** |
| Standard deviation | 0.009 | 0.790 | 1.01 | 0.909 |

**Note.** The confidence threshold is again 0.15.

**Table 5**
YOLOv11m Model $k$-fold Cross-validation Results ($k$ = 4) on the Small Image Dataset

| Fold | AP01 | Precision | Recall | F1-score |
|---|---|---|---|---|
| #1 | 0.980 | 99.0% | 96.7% | 97.8% |
| #2 | 0.994 | 99.4% | 98.2% | 98.2% |
| #3 | 0.989 | 98.6% | 97.6% | 98.1% |
| #4 | 0.977 | 98.6% | 96.0% | 97.3% |
| **Mean** | **0.985** | **98.9%** | **97.1%** | **97.9%** |
| Standard deviation | 0.008 | 0.383 | 0.971 | 0.404 |

**Note.** The confidence score threshold is set to 0.15.

mechanism, do not lead to any significant improvement in our specific detection task.

*Comparison between different YOLOv12 model variants.* We evaluated four YOLOv12 model sizes (nano, small, medium, and large) to assess the impact of model capacity on our monoclass detection task. Table 6 presents the results at epoch 160 on the first fold of the small image dataset. While all variants achieve comparable performance, with precision, recall, and F1-score varying by less than 1% across the tested configurations, the medium variant (`m`) emerges as the optimal trade-off. It delivers the highest F1-score (98.1%) while maintaining a reasonable parameter count (20.2M), avoiding the computational overhead of larger models without sacrificing detection quality. Given the marginal gains from scaling beyond medium size, we selected YOLOv12m for subsequent experiments. Avoiding the computational cost of larger models is also important in our context, given the limited access to high-performance computing resources typical of amateur astronomers.

**Table 6**
YOLOv12 Comparison across Different Model Sizes on the Small Image Dataset

| Model Size | Parameters ($M$) | Precision | Recall | F1-score |
|---|---|---|---|---|
| `n` (nano) | 2.6 | 98.0% | **97.3%** | 97.6% |
| `s` (small) | 9.3 | 98.4% | 96.9% | 97.6% |
| **`m` (medium)** | 20.2 | **99.0%** | **97.3%** | **98.1%** |
| `l` (large) | 26.4 | 98.8% | **97.3%** | 98.0% |

**Note.** The confidence score threshold is set to 0.15.

### *4.4. Large-scale Deployment*

We applied our model to a large portion of the Milky Way, covering the full IPHAS and VPHAS+ surveys. Since the surveys' footprints overlap, this represents a ∼3755 deg$^2$ region.

A set of 1 280 × 1 280 images was generated directly on the CDS servers, with fields of view overlapping by at least 10% in all four directions to limit the truncation of objects. Of the 164,858 images examined, YOLO predicted 49,648 bounding boxes with a confidence score greater than or equal to 0.15. By sorting out duplicates (due to image overlaps) and excluding 3084 objects crossmatched with the HASH database (requiring positional discrepancies in both R.A. and decl. to be under 20″), we end up with 35,665 predictions.

These prediction results are currently being analyzed by members of the AIP-PN project. The most promising candidates will be selected for a follow-up program of photometric and spectroscopic observations, to be carried out in the coming months by a larger team of AIP members. Figure 12 illustrates how the combination of additional spectral bands and longer exposure times in amateur astrophotography can reveal PN features that remain undetected in professional surveys such as IPHAS and VPHAS+, thereby providing complementary evidence for confirming the PN nature of candidates.

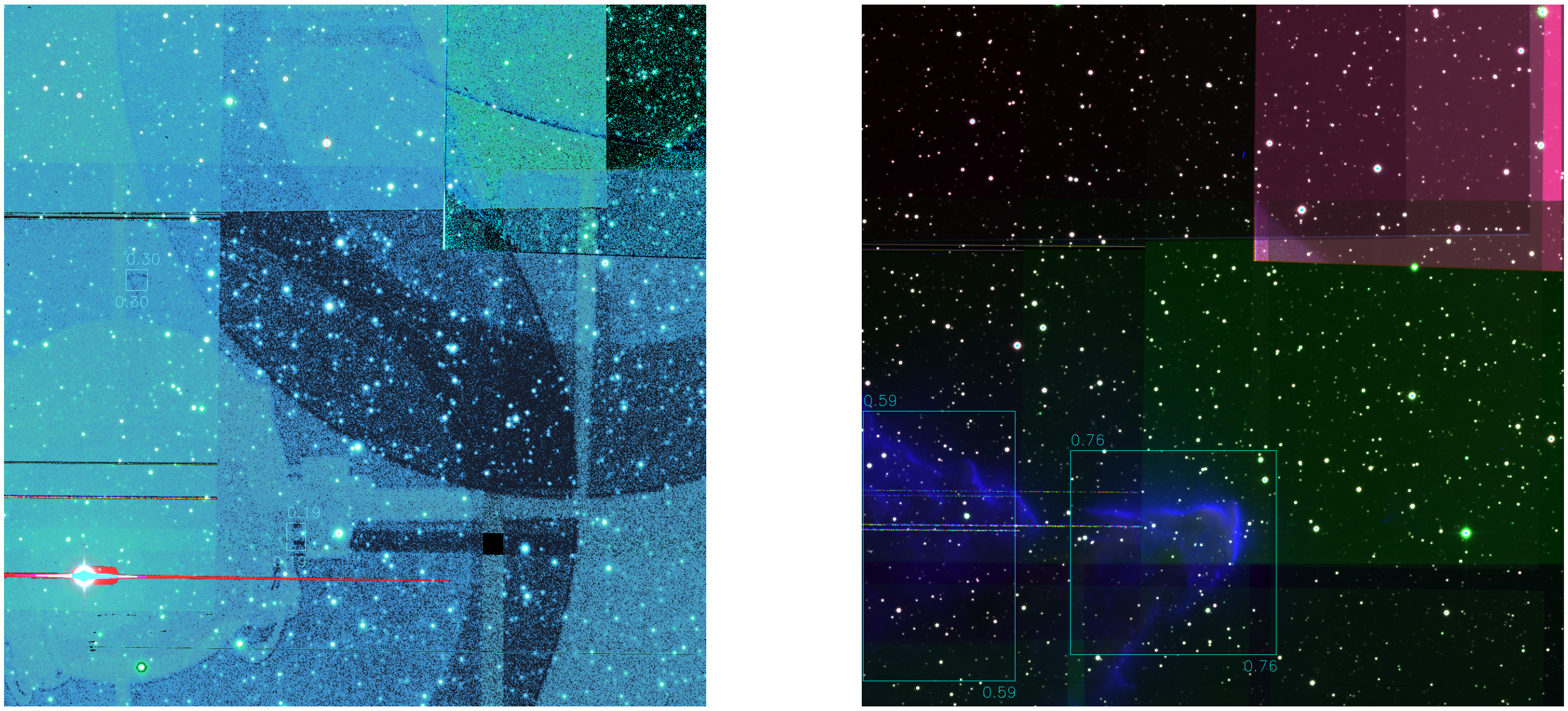

(a) A bright star, probably Gamma Cassiopeiae, triggers several ghost images (increased contrast and brightness).

(b) An extended H II region, called LBN 623, leads to several neighboring false positives with high confidence scores.

**Figure 13.** Examples of common FPs (blue bounding boxes) in IPHAS.

*Common PNe mimics detected by the model.* Predicted contaminants are caused by a variety of PNe mimics and imagery artifacts, including, in particular, some of the CCD issues (see Figure 6), optical issues (ghost images), and most frequently the presence of extended H II regions leading to clusters of neighboring FPs, as shown in Figure 13.

*Comparison with the literature.* In their work, R. Sun et al. (2024) deployed their Swin Transformer model on 43% of the VPHAS+ survey coverage, representing 979 $deg^2$. After analyzing 4.5 million 512 × 512 px images (with a field of view of 107 × 107 $arsec^2$), they obtained ∼20,000 detections. A total of 815 high-quality PNe candidates were identified, among which 31 were prioritized as top-quality targets for optical spectroscopic follow-up. Analysis of their reduced spectra verified that 22 of these objects can be classified as True, Likely, or Possible PNe (Y. Li et al. 2024).

## 5. Conclusion and Prospects

This work demonstrated the excellent performance of YOLO deep learning models in identifying PNe candidates from IPHAS and VPHAS+ astronomical surveys. Combining $r$, $i$, and H$\alpha$ spectral bands in both surveys considered jointly, we produced an augmented dataset of 640 × 640 px high-resolution preprocessed HiPS cutouts and trained our model on examples of True, Likely, and Possible PNe images labeled with single-class bounding boxes based on the information available in the HASH database. The model's loss function was carefully adjusted to account for the single-class use case as well as the diffuse and cloudy morphology of PNe. The model's hyperparameters were experimentally tuned and tracked using MLflow. Convergence occurred after 3.5 hr of training and 160 epochs. After cross validation, we showed that our system achieved an AP (AP01) of 0.979, a precision rate of 97.4%, and a recall rate of 98.0% in the detection of True PNe. These results represent an improvement over those reported in the literature and provide new insights into the comparison between Swin Transformers and YOLO-based deep learning architectures in the PNe detection task.

In addition, we developed a unified, efficient, and versatile pipeline for preprocessing and augmenting astronomical survey data, based on HiPS. Considering the growing number of HiPS available online, we suggest that this methodology could be effectively applied to other deep learning problems in astronomy, such as the detection and classification of other celestial sources in images.

A visual inspection of the predictions of our model shows that it is largely robust to artifacts such as CCD corner defects and HiPS tiling.

To deepen this study, a second dataset of larger images with lower angular resolution was produced. These 1280 × 1280 images have a field of view of 0.04 $deg^2$. The aim was to target more extended PNe, which are generally fainter and rare. Since computational complexity—and consequently model training time—increases with image size, we performed hyperparameter tuning on the small image dataset to avoid the high cost of tuning on the large dataset. However, the model's performance on this dataset is lower, with both precision and recall reaching 94.8%. This degradation may be attributed to the larger field of view, the lower angular resolution of the images, or the increased difficulty of detecting faint, extended PNe.

This second model was then deployed to a large portion of the Milky Way for inference. The analysis of around 34% and 60% of the respective coverages of IPHAS and VPHAS+ surveys led to 21,244 predictions. These results are currently being analyzed by members of the AIP-PN project. The most promising candidates will be selected for a follow-up program of photometric and spectroscopic observations, to be carried out in the coming months by a larger team of AIP members.

Building on the findings presented, this study initiates a discussion on potential directions for future research. The huge number of detections generated by deep learning models when applied to extensive regions of the Milky Way poses a clear challenge. Indeed, a visual analysis of large-scale deployment predictions, as well as predictions on test sets, revealed instances where the model produced clear (to humans) FPs, sometimes with high confidence scores. Though the precision rate is excellent on the test set (which contains a biased proportion of known PNe per solid angle unit), it inevitably drops on large-scale sets due to the rareness of PNe.

Prediction contaminants are caused by a variety of PNe mimics and imagery artifacts, including, in particular, CCD issues, optical issues, and the presence of extended H II regions leading to clusters of neighboring FPs. Therefore, a selection process is necessary before carrying out any follow-up spectroscopic observation. This process currently relies, as a last resort, on human vetting through time-consuming visual inspection. Increasing the confidence threshold is not sufficient and causes the recall rate to fall.

A fundamental challenge persists; the development of an automated system that cannot only reliably detect PNe but also effectively discriminate them from artifacts and, most importantly, from mimics that human experts are able to recognize and distinguish.

To address this problem, we plan to apply a number of improvements to our system. First, we propose annotating a set of background images—including images that were previously misclassified by the model—and incorporating them into the training set, with an appropriate background-to-object ratio. Another approach would be to turn the model into a multiclass detector, with classes corresponding to the most common artifacts and mimics (that can be distinguished on IPHAS and VPHAS+ images by humans).

Another idea is to add extra information to the dataset, or combine our system with other techniques relying on extra information. Future work could focus on utilizing the $g$ band, which encompasses the [O III] spectral line, to investigate the VPHAS+ survey alone. Another possible direction is to apply complementary automated techniques capable of distinguishing PNe candidates from their mimics, such as the infrared-based classification algorithms developed by Akras et al. or the comparison of predictions with the Gaia white dwarf catalog (N. P. Gentile Fusillo et al. 2021) (containing many PNe central stars). It should be noted that such extra data remain incomplete when compared to the established diagnostic process for PNe, which ultimately requires confirmation through individual and costly spectroscopic follow-up observations.

## Acknowledgments

T.S.'s work on this project was funded by a grant from the UHA. A 40 GB GPU was provided by Centrale Méditerranée to train our models. A special thank you to Pascal Le Dû for sharing his experience in PN research with us during these very enriching discussions. T.S. thanks Quentin Parker and Andreas Ritter for agreeing to meet with him at the Strasbourg astronomical observatory to discuss their work, and Sébastien Derriere, who made this meeting possible. Thank you, Jean-Baptiste Marquette, for your technical tips. T.S. thanks Simon Schabat and Julie Munsch for all the interesting discussions they had at IRIMAS about YOLO and machine learning in general.

## ORCID iDs

Titouan Sérandour https://orcid.org/0009-0007-3912-0674
Anne-Laure Méalier https://orcid.org/0009-0008-2686-4006
Bruno Colicchio https://orcid.org/0000-0002-1742-8631
Thomas Boch https://orcid.org/0000-0001-5818-2781
Djaffar Ould Abdeslam https://orcid.org/0000-0001-8900-4566